\newcommand{\beq}{\begin{equation}}
\newcommand{\eeq}{\end{equation}}
\newcommand{\ba}{\begin{array}}
\newcommand{\ea}{\end{array}}
\newcommand{\bean}{\begin{eqnarray*}}
\newcommand{\eean}{\end{eqnarray}}
\newcommand{\bea}{\begin{eqnarray}}
\newcommand{\eea}{\end{eqnarray}}
\newcommand{\bc}{\begin{center}}
\newcommand{\ec}{\end{center}}
\newcommand{\bt}{\begin{table}}
\newcommand{\et}{\end{table}}

\newcommand{\bpsi}{{\bf \psi}}
\newcommand{\la}[1]{\label{#1}}

\newcommand{\ds}{\displaystyle}

\newcommand{\beqno}{\begin{displaymath}}
\newcommand{\eeqno}{\end{displaymath}}

\newcommand{\been}{\begin{enumerate}}
\newcommand{\een}{\end{enumerate}}

\def\ds{\displaystyle}

\documentclass{ws-mplb}

\usepackage{epsfig}

\begin{document}

\markboth{P. G.\ Kevrekidis, R.\ Carretero-Gonz\'alez, D. J.\ Frantzeskakis, and I. G.\ Kevrekidis}{VORTICES IN BOSE-EINSTEIN CONDENSATES}

%
\catchline{}{}{}{}{}
%

\title{Vortices in Bose-Einstein Condensates: Some Recent Developments
}

\author{P. G.\ KEVREKIDIS}

\address{Department of Mathematics and Statistics, 
University of Massachusetts, \\
Amherst, MA 01003, USA \\
kevrekid@math.umass.edu}

\author{R.\ CARRETERO-GONZ\'ALEZ}

\address{Nonlinear Dynamical Systems Group\footnote{URL: http://nlds.sdsu.edu/}, \\
Department of Mathematics and Statistics, \\
San Diego State University, San Diego CA, 92182-7720, USA, \\
http://www.rohan.sdsu.edu/$\sim$rcarrete}

\author{D. J.\ FRANTZESKAKIS}

\address{Department of Physics, University of Athens, 
Panepistimiopolis, \\
Zografos, Athens 15784, Greece\\
dfrantz@cc.uoa.gr}

\author{I. G.\ KEVREKIDIS}

\address{Department of Chemical Engineering and PACM, \\
Princeton University, \\
6 Olden Str.\ Princeton, NJ 08544 USA, \\
yannis@princeton.edu}

\maketitle

\begin{history}
\received{10/30/04, Modern Physics Letters B in Press 2005}
\end{history}

\begin{abstract}
In this brief review we summarize a number of recent developments
in the study of vortices in Bose-Einstein condensates, a topic of
considerable theoretical and experimental interest in the past few
years. We examine the generation of vortices by means of phase
imprinting, as well as via dynamical instabilities. Their stability
is subsequently examined in the presence of purely magnetic trapping, 
and in the combined presence of magnetic and optical trapping.
We then study pairs of vortices and their interactions, illustrating
a reduced description in terms of ordinary differential equations for
the vortex centers. In the realm of two vortices we also consider
the existence of stable dipole clusters for 
two-component condensates. Last but not least, we discuss 
mesoscopic patterns formed by 
vortices, the so-called vortex lattices and analyze some of their 
intriguing dynamical features. A number of interesting future directions 
are highlighted.
\end{abstract}

\section{Abbreviations}


\begin{tabular}{ll}
$\circ$ BEC: Bose-Einstein condensate &
$\circ$ OL: Optical lattice \\[0.0ex]
$\circ$ GP: Gross-Pitaevskii (Equation) &
$\circ$ PR: Parrinello-Rahman \\[0.0ex]
$\circ$ MD: Molecular dynamics &
$\circ$ TF: Thomas-Fermi \\[0.0ex]
$\circ$ MT: Magnetic trap &
$\circ$ NLS: Nonlinear Schr{\"o}dinger (Equation) ~~~
\end{tabular}

\section{Overview}

Bose-Einstein condensation (BEC) was theoretically predicted by Bose and Einstein in 
1924\cite{bose,einstein1,einstein2}\footnote{We should mention that 
part of this overview section has 
significant overlap with the earlier review of two of the present 
authors on ``Pattern Forming 
Dynamical Instabilities of 
Bose-Einstein Condensates'' (Ref. [51] herein); it is
included, however, here for reasons of completeness.}. It consists of the macroscopic occupation of the ground state of a gas of bosons, below a critical transition temperature $T_c$, i.e., a quantum phase transition. However, this prediction was only experimentally verified after 70 years, by an amazing series of experiments in 1995 in dilute atomic vapors, namely of rubidium\cite{anderson} and sodium\cite{davis}. In the same year, first signatures of the occurrence of BEC were also reported in vapors of lithium\cite{bradley} (and were later more systematically confirmed). The ability 
to controllably cool alkali atoms 
(currently over 35 groups around the world can routinely produce BECs) at 
sufficiently low temperatures
and confine them via a combination of magnetic and optical techniques (for a review see e.g., Ref.\ [\refcite{review}]), has been instrumental in this major feat whose significance has already been acknowledged through the 2001 Nobel prize in Physics.

This development is of particular interest also from a theoretical/mathematical standpoint. 
On the one hand, there is a detailed 
experimental control over the produced BECs. On the other, equally importantly, there is a very good model partial differential equation (PDE) that can describe, at the mean field level, the behavior of the condensates.
This model (which, at heart, approximates a quantum many-body interaction with a classical, but nonlinear self-interaction) is the well-known Gross-Pitaevskii 
equation\cite{gross,pittaevskii}, a variant of the famous Nonlinear Schr{\"o}dinger equation (NLS)\cite{sulem} that reads:
\begin{equation}
i \hbar \Psi_t=-\frac{{\hbar}^2}{2 m} \Delta \Psi + g |\Psi|^2 \Psi + 
V_{{\rm ext}} ({\bf r}) \Psi,
\label{peq1}
\end{equation}
where $\Psi$ is the mean-field condensate wavefunction (the atom density is 
$n=|\Psi(x,t)|^2$), 
$\Delta$ is the Laplacian, $m$ is the atomic mass, and the nonlinearity coefficient 
$g$ (arising from the interatomic interactions) is proportional to the atomic scattering length\cite{review}. This coefficient is positive (e.g., for rubidium and sodium) or negative (e.g., for lithium) 
for repulsive or attractive interatomic 
interactions respectively, 
corresponding to defocusing or focusing 
cubic (Kerr) nonlinearities in the context 
of nonlinear optics\cite{sulem}. 
Notice, however, that
experimental ``wizardry'' can even manipulate the scattering length (and, thus, the nonlinearity coefficient $g$) using the so-called Feshbach resonances\cite{feschb} to achieve
any positive or negative value of the scattering length at will (i.e., 
nonlinearity strength in Eq.\ (\ref{peq1})). Moreover, the external potential $V_{{\rm ext}}$ can assume different forms. For the ``standard'' magnetic trap (MT) usually implemented to confine the condensate, 
this potential has a typical harmonic form:
\begin{equation}
V_{{\rm MT}}=\frac{1}{2} m \left( \omega_x^2 x^2 + \omega_y^2 y^2 + \omega_z^2 z^2\right),
\label{peq2}
\end{equation}
where, in general, the trap frequencies $\omega_x, \omega_y, \omega_z$ 
along the three directions are different. As a result, in recent experiments, the shape of the trap (and, hence, the form of the condensate itself) can range from isotropic to the so-called cigar-shaped traps (see, e.g., Ref.\ [\refcite{review}]), to quasi-two-dimensional\cite{catal1}, or even quasi-one dimensional forms\cite{catal2}. 
Moreover, linear ramps of (gravitational) potential $V_{{\rm ext}} = m g z$ have also been experimentally used\cite{kasevich}. Another prominent example of an experimentally
feasible  potential is imposed by a pair of laser beams forming a standing wave which generates a periodic optical potential, the so-called optical lattice (OL)\cite{ol1,ol2,tromb,konot},
of the form:
\begin{equation}
V_{\rm OL}=V_0 \left[\cos^2\left(\frac{2 \pi x}{\lambda_{x}} + \phi_x\right)
                   + \cos^2\left(\frac{2 \pi y}{\lambda_{y}} + \phi_y\right)
                   + \cos^2\left(\frac{2 \pi z}{\lambda_{z}} + \phi_z\right) \right],
\label{peq3}
\end{equation}
where $\lambda_{x,y,z}=\lambda_{\rm laser} \sin(\theta_{x,y,z}/2) /2$, $\lambda_{\rm laser}$ is the laser wavelength, $\theta$ is the angle between the laser beams\cite{morsch}, and $\phi_{x,y,z}$ is a phase detuning factor (both the latter are potentially variable). Such potentials 
have been realized in one\cite{ol1,ol2}, two (the so-called egg-carton 
potential)\cite{haensch} and three dimensions\cite{catal1,greiner}. 

Moreover, present experimental realizations render feasible/controllable the adiabatic or abrupt displacement of the magnetic or optical lattice trap\cite{morsch,kas2,catal3} 
(inducing motion of the condensates), the ``stirring'' of the condensates providing angular momentum and creating excitations with topological charge such as 
vortices\cite{vort1,vort2,vort3} and vortex lattices 
thereof\cite{latt1,latt2,latt3}. Additionally, phase engineering of the condensates is also feasible experimentally\cite{denschl}, and this technique has been used in order to produce nonlinear matter-waves, such as dark 
solitons\cite{dark1,dark2} in repulsive BECs. Note that, more recently, bright solitons have been generated as well\cite{bright1,bright2} in attractive BECs, and both types are currently being studied extensively.

Among these coherent structures, of particular interest are the nonlinear waves of non-vanishing vorticity. Vortices are worth studying not only due to their significance as a fundamental type of coherent nonlinear excitations but also because they play a dominant role in the breakdown of superflow in Bose fluids\cite{jac11,jac12,jac13}. The theoretical description of vortices in BECs can be carried out in a much more efficient way than in liquid He (see Ref.\ [\refcite{jac1}]) due to the weakness of the interactions in the former case. 
These advantages explain a large volume of work regarding the behavior of vortices in BECs, some of which has been 
summarized in Ref.\ [\refcite{fetter}]. It is interesting to note in this connection that the description of such topologically charged nonlinear waves and their surprisingly ordered and
robust lattices, as well as their role in phenomena as rich and profound as superconductivity and superfluidity were connected to 
the theme of the recent Nobel prize in Physics in 2003.

It is around this exciting frontier of theoretical and experimental studies between atomic physics, soft condensed-matter physics and nonlinear dynamics that this brief review is going to revolve. Our aim is to report some recent developments in the study of vortices in the context of mean field theory (i.e., employing the GP equation). We consider various external potentials relevant to the trapping of the condensates, examining the existence, generation and dynamical stability of such coherent structures. It should be noted, however, that all the perturbations considered herein will preserve the Hamiltonian structure of the system. We will 
discuss these notions at various levels of increasing complexity, starting from that of a single vortex (in Section \ref{sec:SingleVortex}), proceeding to that of a few vortices and using the two-vortex system as a characteristic example (in Section \ref{sec:TwoVortices}), while in Section \ref{sec:VortexLattices}, we will address the behavior of large clusters of vortices, the so-called vortex lattices. Section \ref{sec:Conclusions} summarizes our findings and presents some interesting directions for future studies. 

\section{Single Vortex}\label{sec:SingleVortex}

\subsection{Generation}\label{subsec:Generation}

It is well-known that if a superfluid is subjected to rotational motion, 
vortices will be generated in it.
Such a situation also occurs in dilute BECs, where quantized vortices can be described in the framework of the GP equation. Thus, in this context, a vortex can typically be created upon ``stirring''
the condensate. In particular, beyond a critical angular velocity $\Omega_c$, the
energy functional associated with the GP equation (incorporating the centrifugal term due to rotation), namely
\begin{equation}
E=\int d  {\bf r} \left[ -\frac{1}{2} \Psi^{\star} \Delta \Psi + 
V_{\rm ext} |\Psi|^2 
+ \frac{g}{2} |\Psi|^4 - \Omega_z \Psi^{\star} L_z \Psi \right],
\label{freeE}
\end{equation}
is minimized by a single vortex configuration\cite{perez}, 
resulting in the generation of such a structure, 
observed also experimentally\cite{vort2}.
Note that in Eq.\ (\ref{freeE}), $\Omega_z>\Omega_c$ is the angular velocity and 
$L_z=i(x\partial_y-y\partial_x)$ represents the angular momentum along the $z$ axis ($\star$ stands for complex conjugate).

However, vortices can be spontaneously produced in a number of alternative ways, some of which we examine in what follows. More specifically:
\begin{enumerate}
\item they can {\it spontaneously emerge} as the pattern
forming outcome of {\it dynamical instabilities}, or 
\item they may be 
{\it imprinted} on the condensate via appropriate modulation of 
its {\it phase}. 
\end{enumerate}

One instability that can be exploited as a method of producing vortex patterns is the 
so-called transverse or (``snaking'') instability of {\it rectilinear} dark solitons. 
This instability, which has been studied extensively in the context of nonlinear optics (see, e.g., Ref.\ [\refcite{kivpr,kivpel}] for a review), forces a dark-soliton to undergo transverse modulations that cause the nodal plane to decay into vortex pairs. The snake instability is 
known to occur in trapped BECs as well (see Ref.\ [\refcite{dark2}] for its experimental observation and Ref.\ [\refcite{shlyap1}] for relevant analytical and numerical results). In this context, dark solitons are placed on top of an inhomogeneous background, the so-called Thomas-Fermi (TF) cloud, which approximately yields the ground state wavefunction in the case of repulsive condensates 
($g>0$)\cite{review}, and can be expressed as
\begin{equation}
\Psi_{\rm TF}=\exp(-i \mu t) \sqrt{\frac{ {\rm max}\left\{\mu-V_{\rm ext},0\right\}}{g}},
\label{tfcloud}
\end{equation}
where $\mu$ is the condensate's chemical potential. The snake instability of dark solitons in trapped BECs sets in whenever the soliton motion is subject to a strong coupling between the longitudinal and transverse degrees of freedom, i.e., far from 1D geometries. A relevant discussion demonstrating how the snaking instability manifests itself as the transverse confinement becomes weak, giving rise to the formation of vortices can be found in the recent work
of [\refcite{npp1}]. 

To quantify better the above, we follow Ref.\ [\refcite{andrea}] to analyze the transverse instability via length-scale competition arguments. In particular, we first consider the following dimensionless 2D version (relevant for a quasi-two-dimensional, or ``pancake'' BEC lying on the $x-y$ plane) of the original GP equation, 
\begin{equation}
i \Psi_t=-\frac{1}{2} \Delta_{\perp} \Psi + |\Psi|^2 \Psi + V_{{\rm ext}} (r) \Psi,
\label{ngp}
\end{equation}
in which length is scaled in units of the fluid healing length $\xi=\hbar/\sqrt{n_{0}gm}$ ($n_{0}$ is the peak density of the gas in the radial direction), $t$ in units of $\xi/c$ (where $c=\sqrt{n_{0}g/m}$ is the Bogoliubov speed of sound), and the atomic density is rescaled by the peak density $n_{0}$; finally, the external potential is $V_{\rm ext}(r)=(1/2) \Omega^2 r^2$, where $r^{2}=x^{2}+y^{2}$ and the parameter 
$\Omega = \sqrt{\omega_{\perp}/\omega_{z}}(4\pi a l_{\perp} n_{0})^{-1}$   
(where $l_{\perp}=\sqrt{\hbar / m \omega_{\perp}}$ is the transverse harmonic oscillator length, $\omega_{\perp}$ being the transverse confining frequency) expresses the dimensionless effective magnetic trap strength. In the context of Eq.\ (\ref{ngp}), and in the absence of the potential, the transverse instability occurs  
for perturbation wavenumbers  
\begin{equation}
k < k_{cr} \equiv \left[ 2 \sqrt{\sin^4{\phi}+ 
\cos^2{\phi}} - \left(1+\sin^2(\phi) \right)
\right]^{1/2}, 
\label{perturb}
\end{equation}
where $\cos\phi$ is the dark-soliton amplitude (depth) and $\sin{\phi}$ is its velocity\cite{kivpr}. In the case of stationary (black) solitons $\cos\phi=1$, hence $k_{cr}=1$. On the other hand, in the presence of the potential,
the characteristic length scale of the BEC (i.e., the diameter of the cloud in the TF approximation) is 
$R_{\rm BEC}=2^{3/2} \mu_{0}^{1/2}/\Omega$, where $\mu_{0}$ is the dimensionless chemical potential. Then, one can argue that the criterion for the suppression of the transverse instability is that the scale of the BEC be shorter than the minimal one necessary for the onset of the instability, leading to the condition
%
\begin{equation}
\Omega > \frac{\sqrt{2 \mu_{0}}}{\pi}.
\label{eqn6}
\end{equation}
If inequality (\ref{eqn6}) is not satisfied, the transverse instability should develop, resulting in the breakup of the dipole configurations (resulting from a dark soliton, 
truncated by the Thomas-Fermi state) into vortex-antivortex pairs. 
It is relevant to note that similarly to the case of rectilinear solitons, the snake instability of the {\it ring} dark solitons can also be responsible for the creation of vortices and vortex arrays as well. In particular, as far as the ring dark solitons are concerned, they were previously predicted in the context of nonlinear optics\cite{KYR}, where their properties were studied both theoretically\cite{yuri2} and experimentally\cite{yuri3}. These entities were also found to exist in the context of BECs\cite{rdsprl}
(see also the relevant recent work\cite{nr}). In the latter context, and in the case of sufficiently large initial soliton amplitudes, ring dark solitons were observed to be dynamically unstable towards azimuthal perturbations that led to (snaking and) their breakup into vortex-antivortex pairs, as well as robust vortex arrays, the so-called ``vortex necklaces''. The latter consist of 
four vortex-pair patterns, with their shape alternating between an ``X'' and a cross (for details, see Figs.\ 3 and 4 of Ref.\ [\refcite{rdsprl}]). 
We do not discuss these instabilities further, as they were analyzed in some detail in the recent review of Ref.\ [\refcite{kf}].


\begin{figure}[tbp]
\begin{center}
\epsfig{file=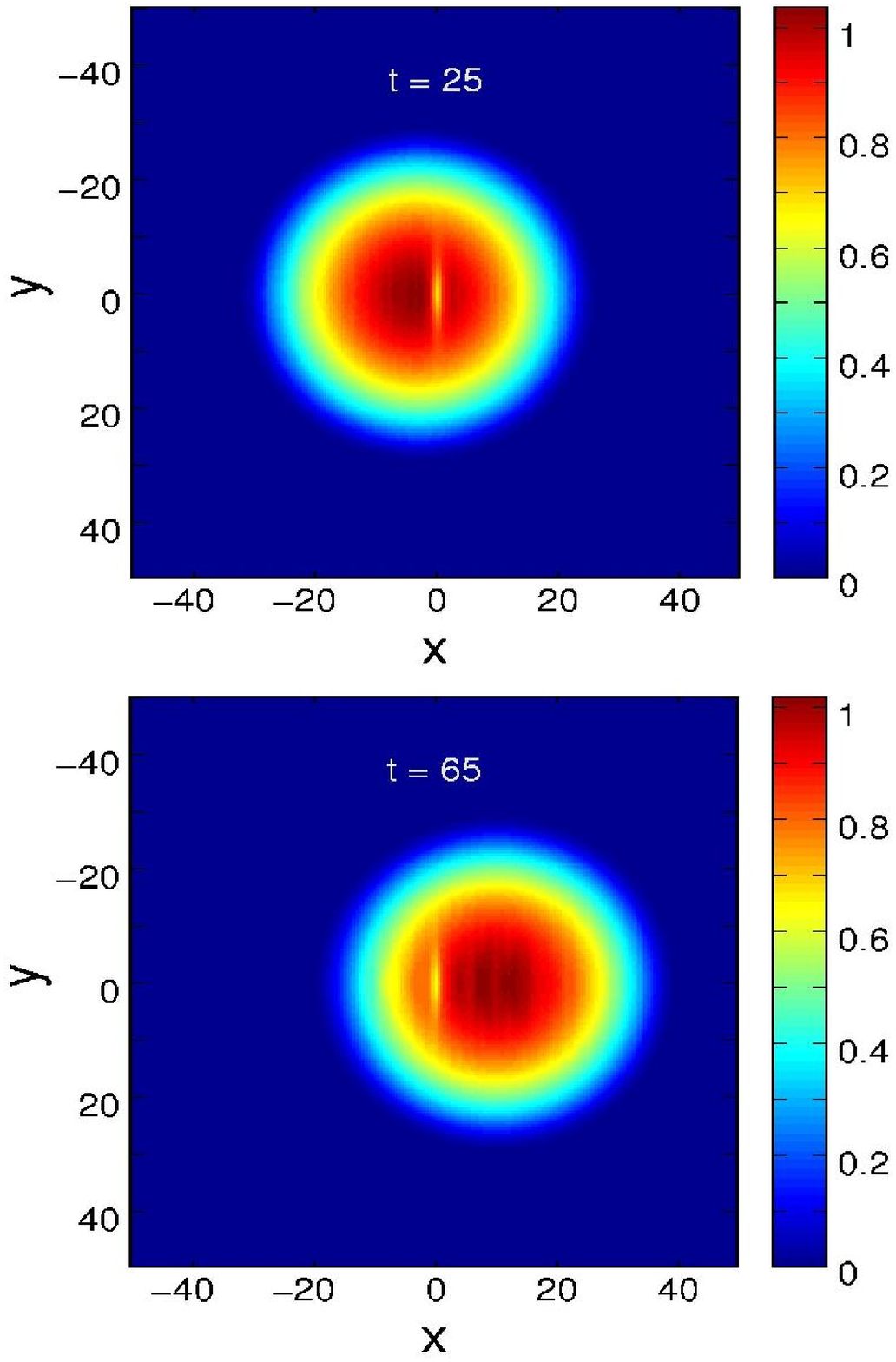, width=1.5in,  height=2.5in, angle=0,silent=}
~~~~~~~
\epsfig{file=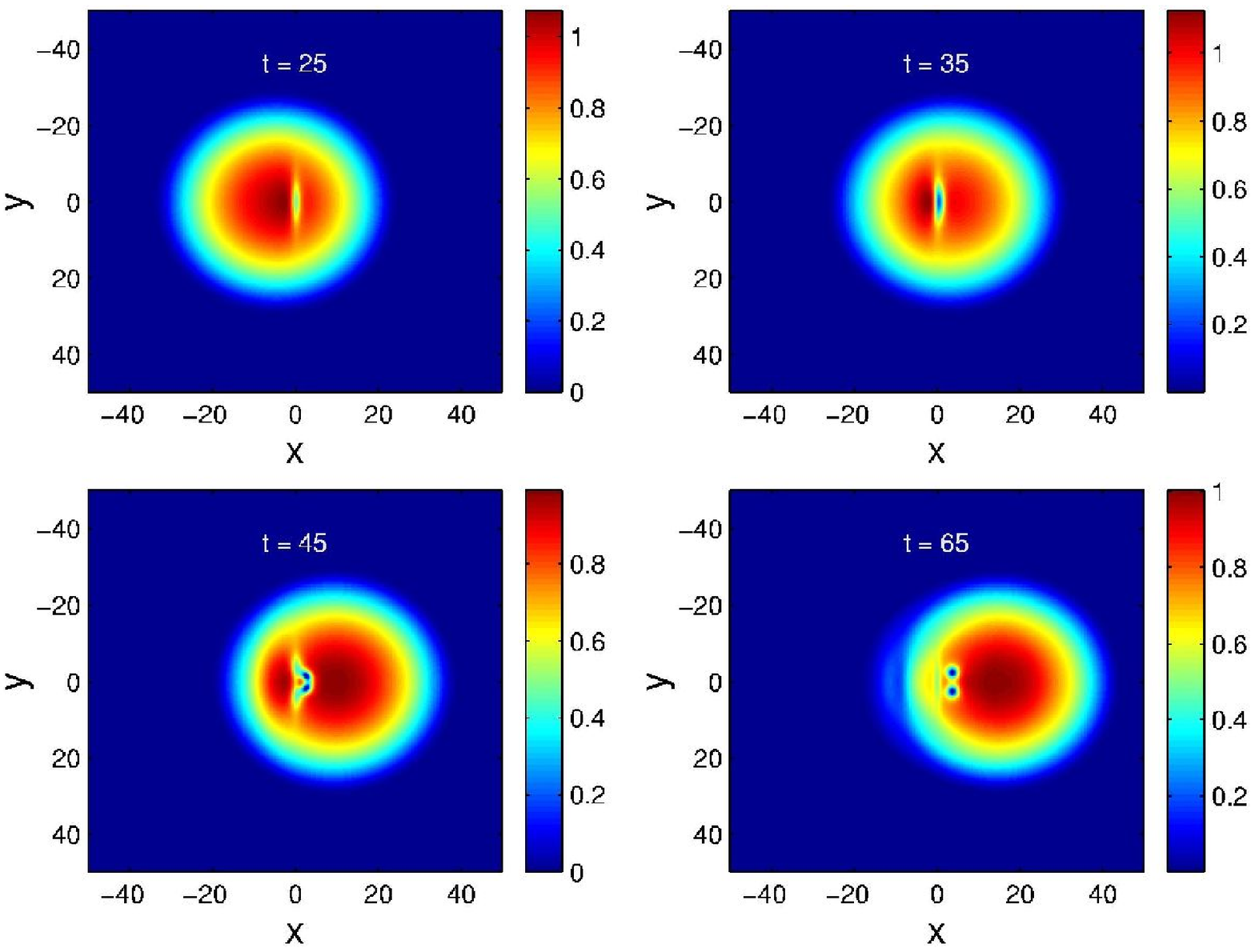, width=3.0in, height=2.5in ,angle=0,silent=}
\end{center}
\caption{Comparison of a subcritical case for $x_0=10$ 
with a supercritical one for $x_0=15$. The 2 subplots on the left 
show different time snapshots of the evolution of the two-dimensional
contour plot of the wavefunction square modulus for
$x_0=10$. The 4 subplots of the 
right show the corresponding snapshots for $x_0=15$
in the unstable case, leading to the emission of a vortex pair.}
\label{mfig1}
\end{figure}


A context similar to that of pattern forming instabilities, and one which may be loosely related to the Landau instability of superflows, involves the interaction of the condensate with an impurity in a recently proposed dynamical experiment\cite{landpaper}
(which bears resemblances to the phenomenon of vortex trailing in fluids, see e.g., Ref.\ [\refcite{saf}]). In particular, in Ref.\ [\refcite{landpaper}] (and in the framework of the dimensionless GP Eq.\ (\ref{ngp})), the magnetic trap originally trapping the condensate at $(0,0)$, was proposed to be displaced by a displacement of $x_0$, but also in the presence of an additional anisotropic impurity potential of the form $V_{\rm imp} = V_0\, {\rm sech}^2\left(\sqrt{(x/r_x)^2+(y/r_y)^2}\right)$ with $V_0=0.5$, $r_y=10$ and $r_x=5$. If the displacement (which was giving rise to an oscillating motion of the condensate) was subcritical, then it was observed that it did not lead to the formation of ``coherent structure radiation''. For supercritical displacements, it was observed to give rise to the formation of vortex pairs, as shown in the results of Fig.\ \ref{mfig1}. The critical speed (proportional to the critical displacement) acquired by the condensate upon ``impact'' with the impurity was numerically observed in Ref.\ [\refcite{landpaper}] to closely match an effective speed of sound in the inhomogeneous medium (i.e., in the presence of the parabolic potential), indicating the possibility to attribute the vortex production in this context
to a Landau-type instability\cite{landau,wuniu2}.

\begin{figure}[th]
\centerline{
\epsfig{figure=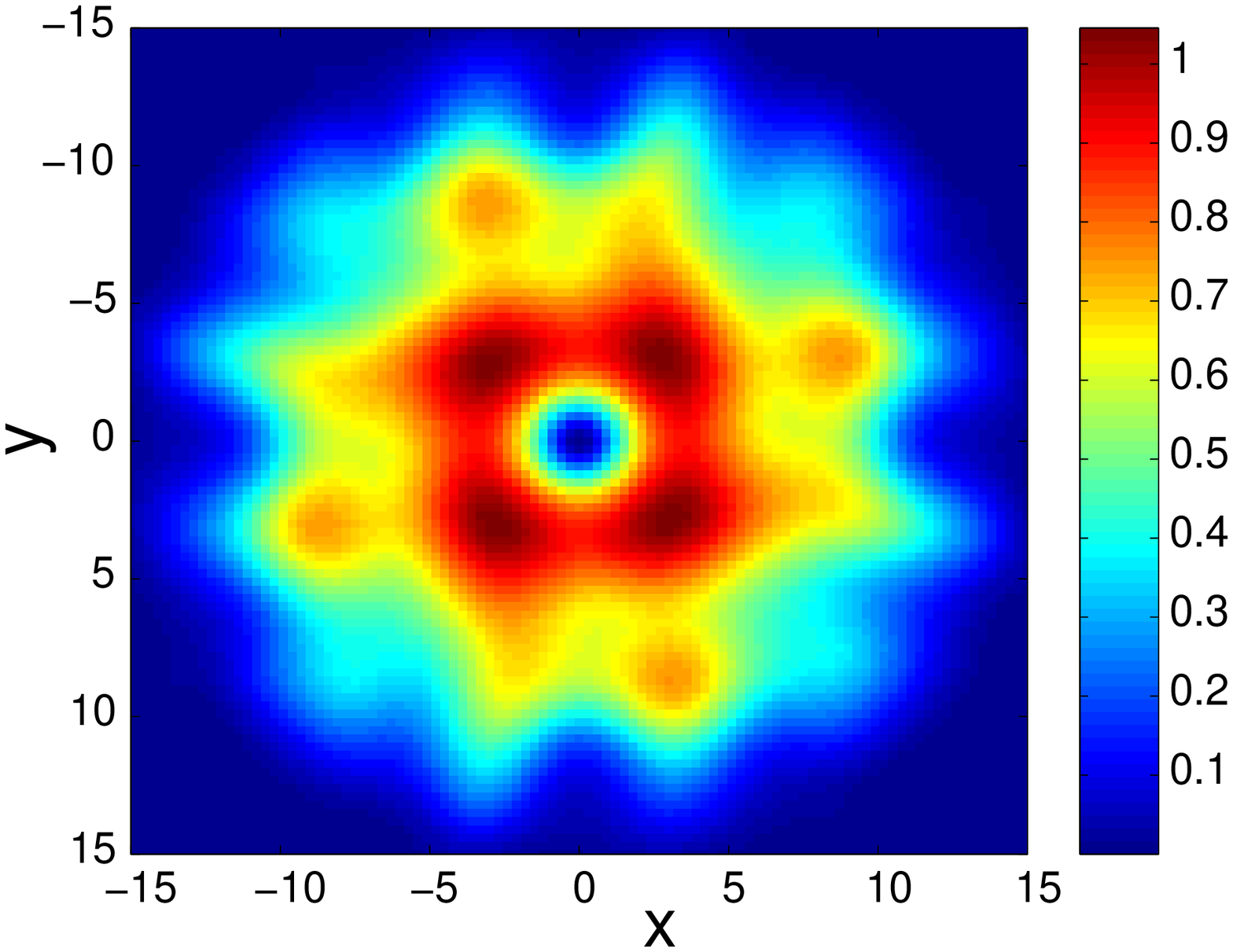,height=1.23in,angle=0,silent=}~~
\epsfig{figure=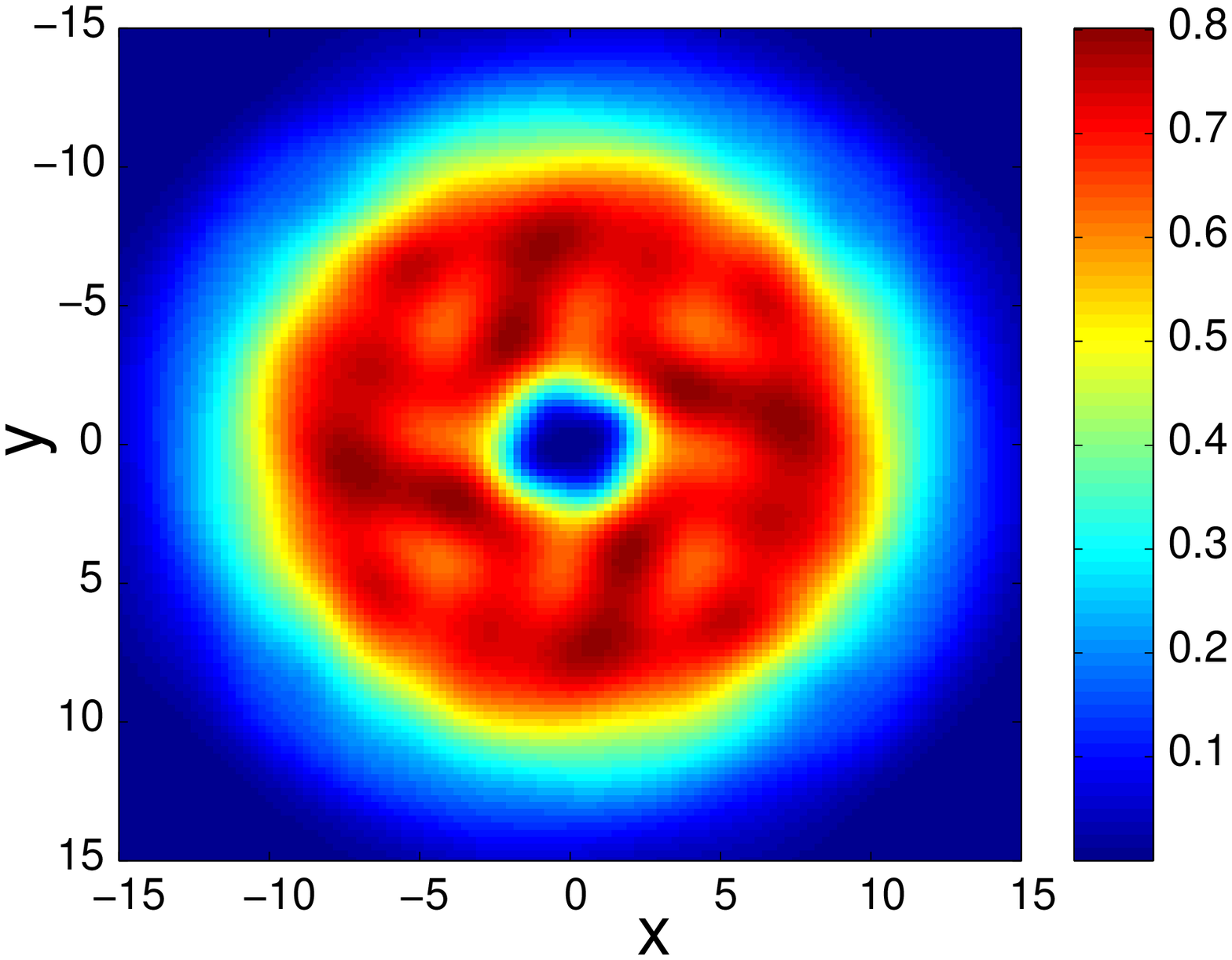,height=1.23in,angle=0,silent=}~~
\epsfig{figure=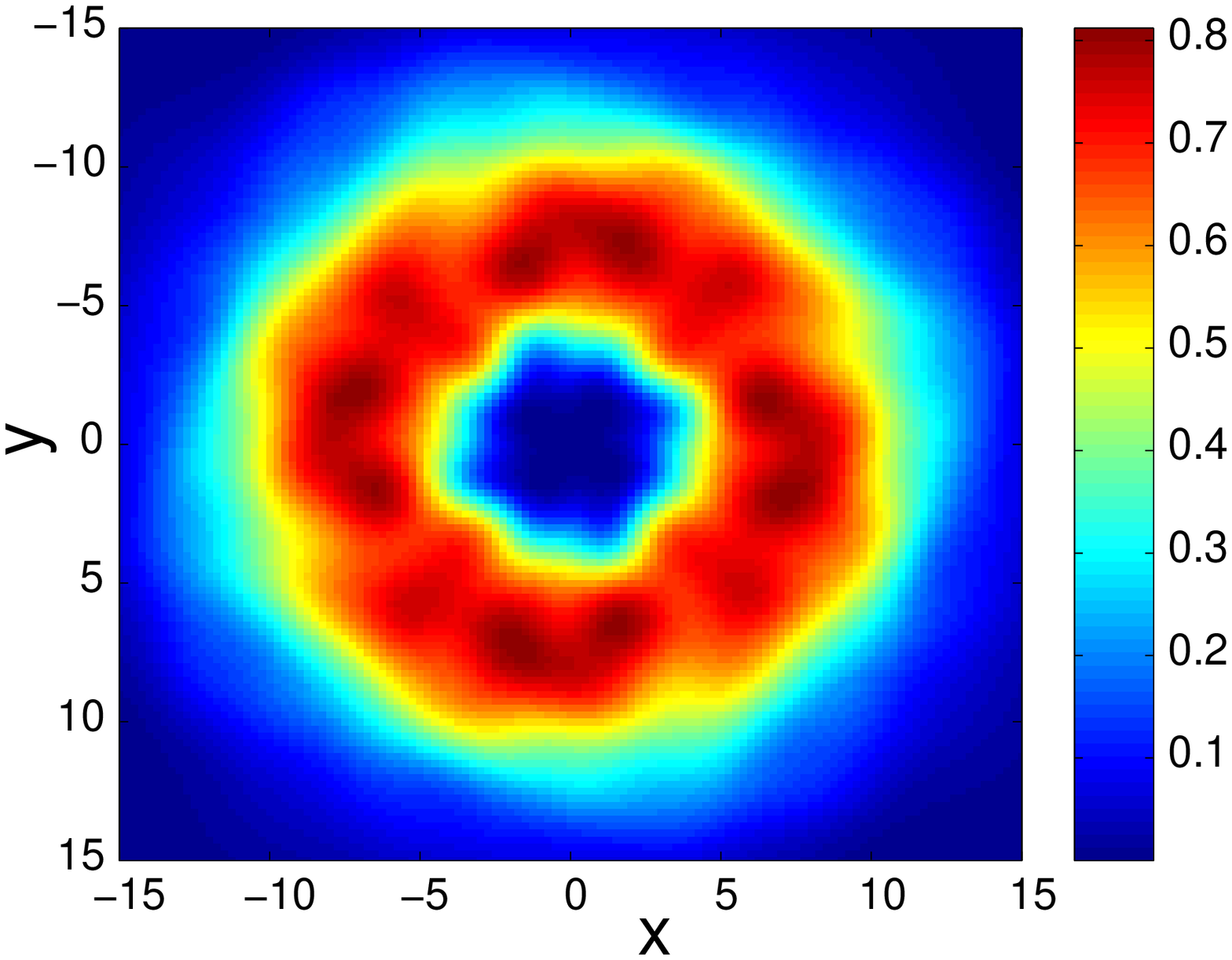,height=1.23in,angle=0,silent=}
}
\vspace*{8pt}
\centerline{
\epsfig{figure=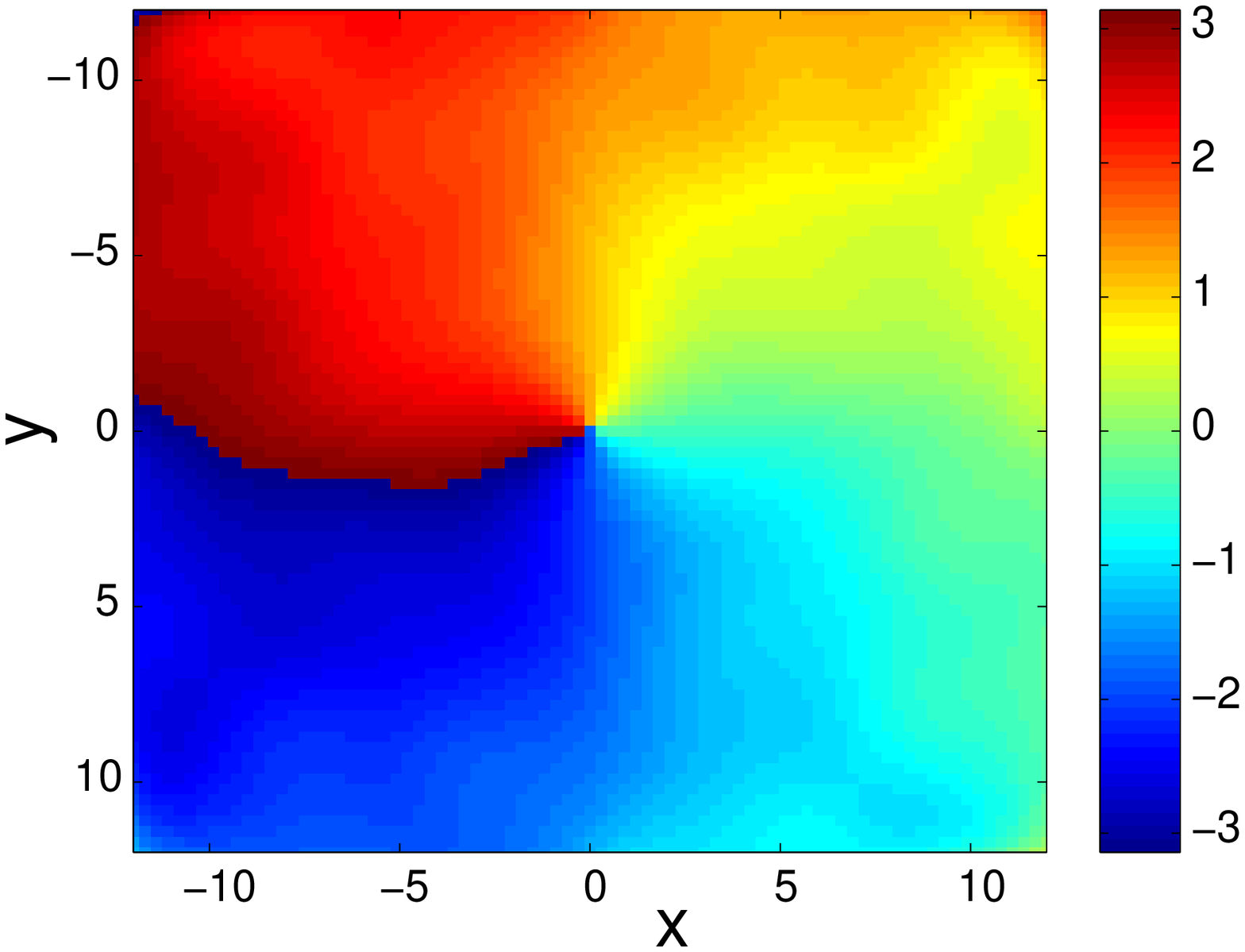,height=1.23in,angle=0,silent=}~~
\epsfig{figure=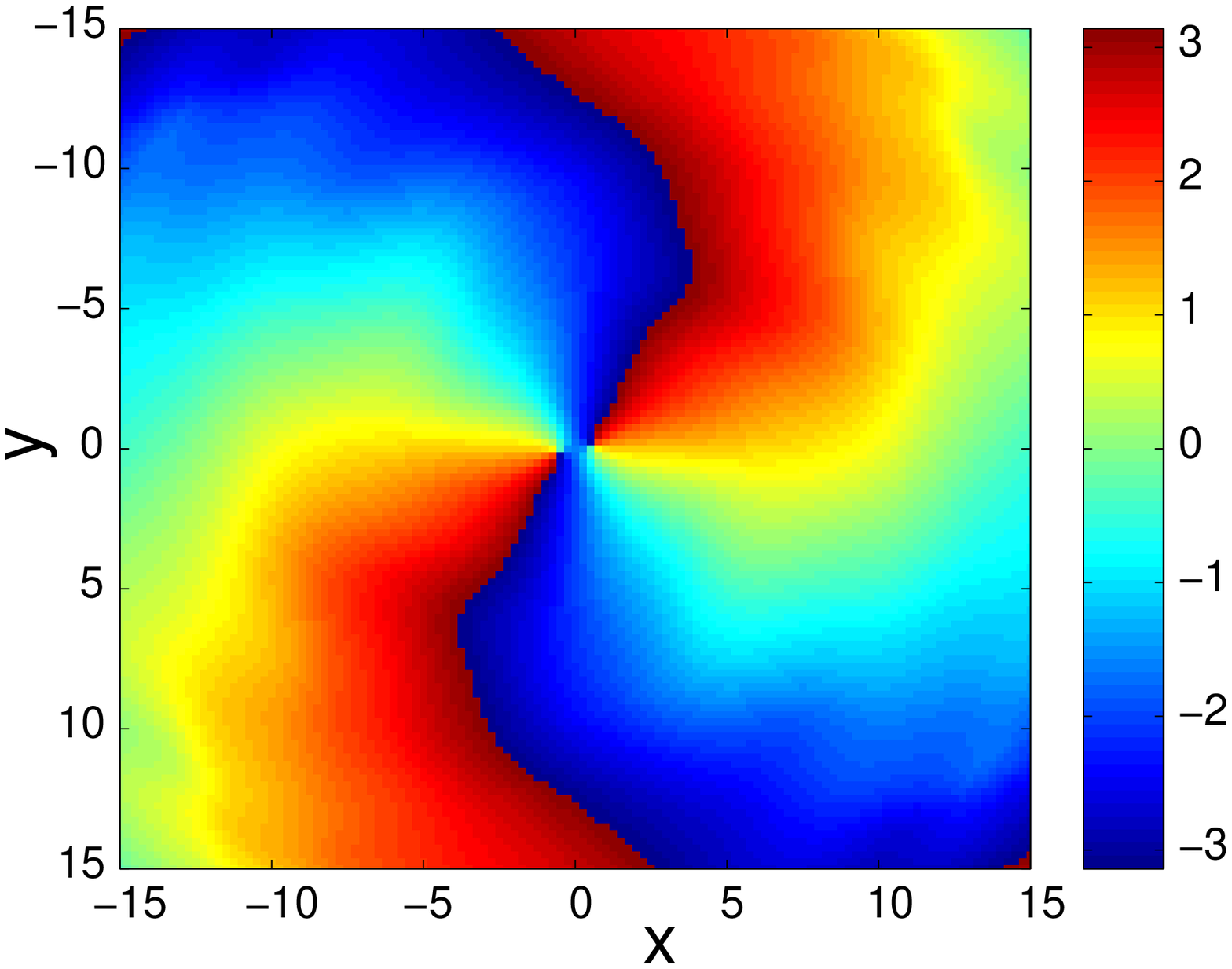,height=1.23in,angle=0,silent=}~~
\epsfig{figure=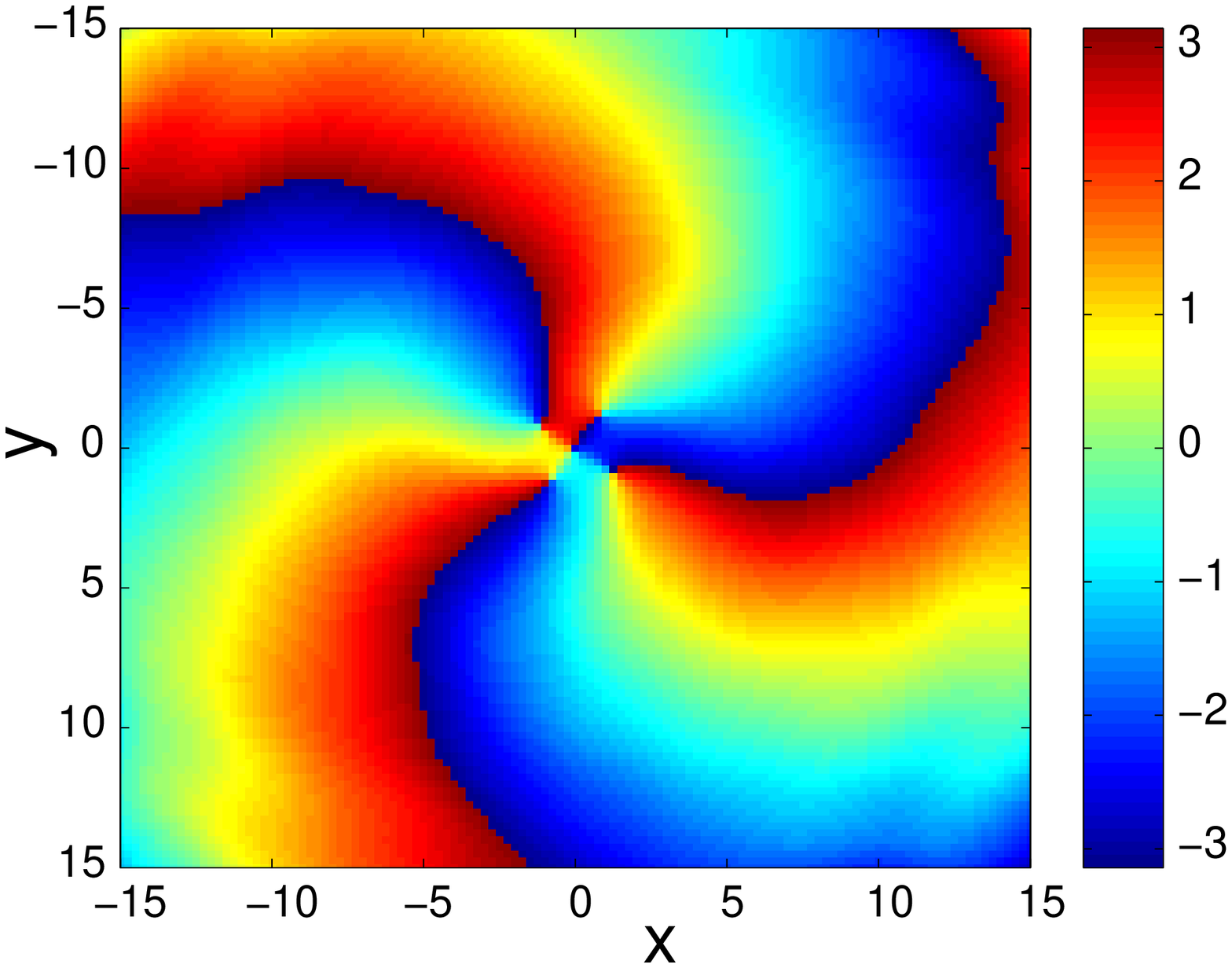,height=1.23in,angle=0,silent=}
}
\caption{Phase imprinting of vortices inside the combined
magnetic and optical trap (left for vortices of charge
$S=1$) and solely in the magnetic trap (middle and right
for $S=2$ and $S=3$ respectively). The top and bottom row 
depict, respectively, the atomic density and phase. Left to
right: vortices of increasing topological charge of one, 
two and three. In all cases, a radial magnetic trap frequency
of $\Omega=0.1$ has been used, while for the left panel
the optical has been chosen with $\lambda_x=\lambda_y=4 \pi$,
and $\phi_x=\phi_y=0$, while $V_0=0.25$.}
\label{mfig2-3}
\end{figure}

Let us conclude this section,by 
briefly discussing one of the standard techniques that have been implemented to create dark solitons and vortices in trapped BECs, namely the so-called phase imprinting or phase engineering technique. According to the latter, a dark soliton can be  generated upon imprinting a phase difference of $\pi$ along the 
condensate\cite{denschl,dark1,dark2}. On the other hand, in the case of vortices, imprinting (through an appropriate ``phase mask'') of a phase difference of $2 \pi$ around a contour can generate  vortex structures (which carry topological charge). In fact, this method was proposed as a means of preparing 
vortex states in 2D BECs in Ref.\ [\refcite{Williams:99}]
and was subsequently used in the laboratory in the experiments of Ref.\ [\refcite{vort1}].
Some of the advantages that this technique has over the previous
ones are that:
\begin{enumerate}
\item It is very robust 
in the presence of even strong perturbations (i.e.,
it works equally well in the cases of combined potentials such as
magnetic and optical ones), see left panels in Fig.\ \ref{mfig2-3}.
\item It can be straightforwardly generalized to produce vortices
of higher charge. 
%
These vortices may also be observed (for appropriate parameter
values) to be dynamically stable for very long times and hence
should, in principle, be experimentally observable; see middle
and right panels in Fig.\ \ref{mfig2-3}.

\end{enumerate}
\subsection{Dynamical Stability}

\subsubsection{Continuum Models}


Let us now address the question of stability of vortices trapped in a combined MT and OL as described in Ref.\ [\refcite{jpbpgk}]. Consider a quasi-two-dimensional\cite{catal1} condensate where a single vortex has been generated at the center of the combined potential
\begin{equation}
V_{{\rm ext}}(x,y) = V_{{\rm MT}}(x,y) + V_{{\rm OL}}(x,y),  
\label{MT+OL}
\end{equation}
where the contributions to the MT and OL, are given by the 2D equivalents
of (\ref{peq2}) and (\ref{peq3}), respectively.
As mentioned in the previous Section, the effective 2D GP equation [cf.\ Eq.\ (\ref{ngp})] applies to situations where 
the condensate has a nearly planar (``pancake'') 
shape, see for example Ref.\ [\refcite{GPE2d}] and references therein. 
Accordingly, vortex states considered below are not subject to 3D 
instabilities (corrugation of the vortex axis\cite{fetter}) as the 
transverse dimension is effectively suppressed.

\begin{figure}[th]
\centerline{
\epsfig{figure=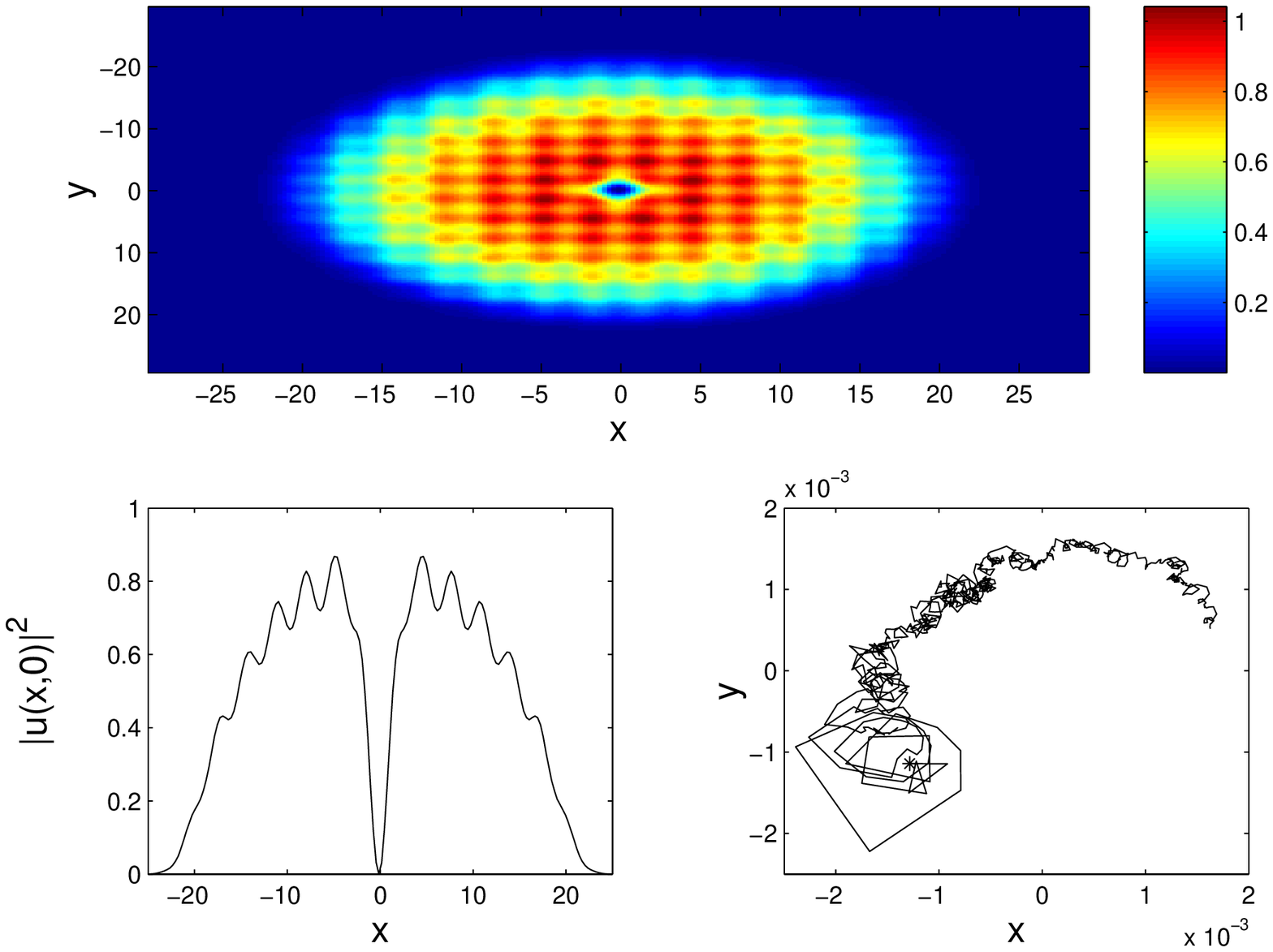,width=6.0cm,angle=0,silent=}
~~~
\epsfig{figure=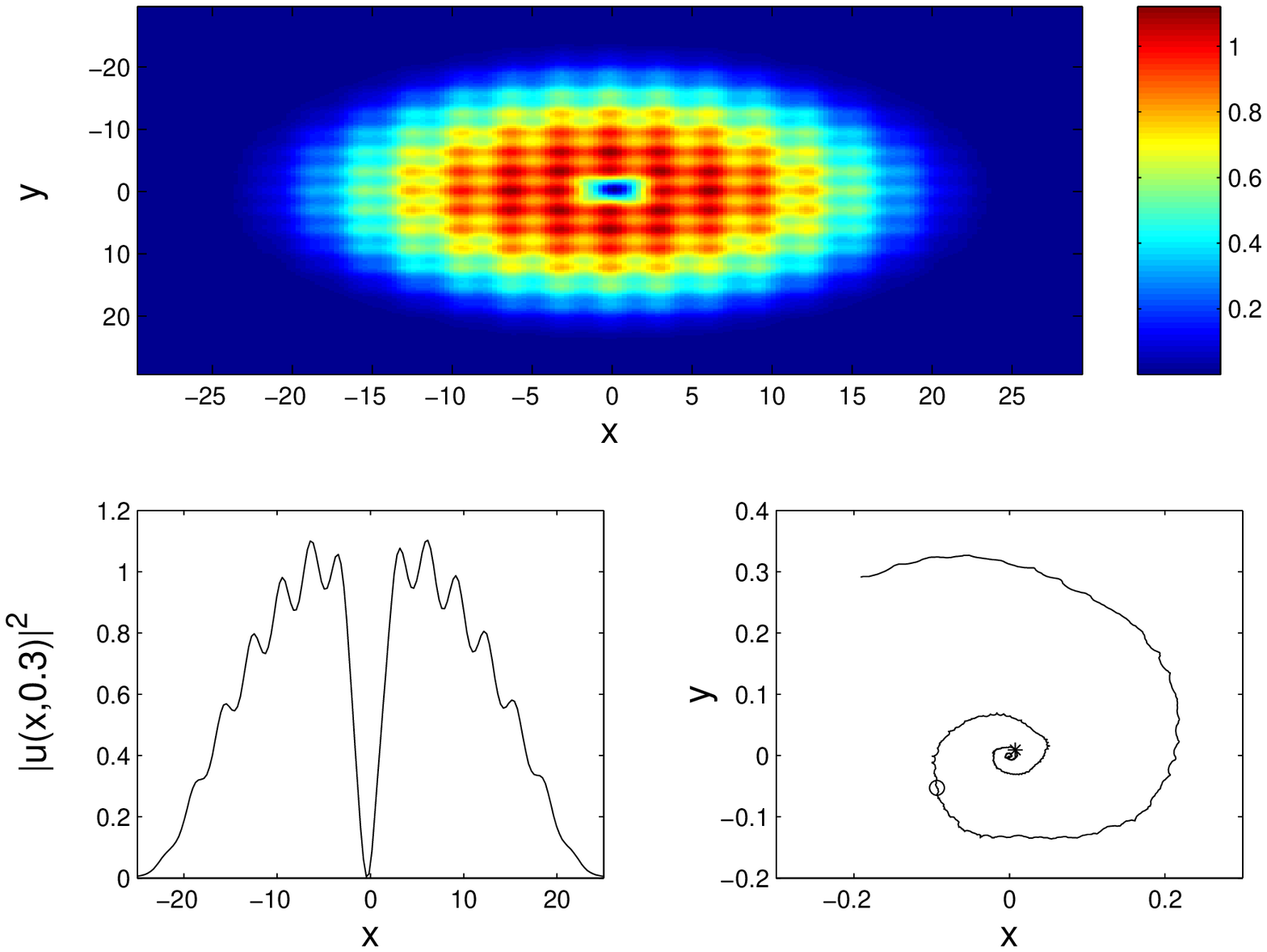,width=6.0cm,angle=0,silent=}
}
\caption{
Vortex stability inside a combined magnetic and optical trap.
The MT and OL parameters are $\omega_x^2=\omega_y^2=0.002$,
$V_0=0.5$ and $\lambda=\lambda_x=\lambda_y=2\pi$.
Left: the vortex is stable at the bottom of a
cosinusoidal OL [$\phi =0$ in Eq.\ (\ref {peq3})].
The top panel shows the contour plot of the
density at $t=100$ ($138$ ms). The bottom left panel is a cut of
the same density profile along $x=0$, while the bottom right one shows the
motion of the vortex center for $0\leq t\leq 100$, the initial position
being marked by a star. Notice the scale ($10^{-3}$, or $1$ nm in physical units)
of very weak motion
of the vortex, which thus stays practically immobile at the origin.
Right: same as left panels but with a
sinusoidal OL [$\phi =\protect\pi/2$] and for a
larger time of $t=250$ ($346$ ms). The bottom right panel
depicts the motion of the vortex center for $0\leq t\leq 250$ [positions of
the vortex center at $t=100$ ($138$ ms) and $t=200$ ($276$ ms),
respectively, are indicated by the star and circle].
}
\label{vortstab}
\end{figure}

The stability of the vortex at the center of the trap can be qualitatively understood in terms of an effective potential obtained by means of a variational approximation\cite{Progress}. As in Ref.\  [\refcite{tempere}], we use the following ansatz to approximate the position 
$r_0(t)=(x_0(t),y_0(t))$ of vortex near the trap center
\begin{equation}
\Psi(x,y,t)=B(t)\,||r_{0}(t)||\exp \left[ -||r_{0}(t)||^{2}/b(t)\right] \,\,
e^{i\varphi_{0}(t)},  \label{veq6}
\end{equation}
where $\varphi_{0}(t)\equiv \tan ^{-1}[(y-y_{0}(t))/(x-x_{0}(t))]$
and $||r_0(t)||$ denotes the 2-norm of the vector $r_0$.
Similarly to the calculations in Refs.\ [\refcite{ricardo1,ricardo2}],
or 
upon employing the conservation of norm, it is straightforward to show that, to leading order, $B(t)=B(0)$ and $b(t)=b(0)$ are approximately constant.
Then, assuming the same detuning and periodicity in all directions 
(i.e., $\phi_x=\phi_y=\phi$ and $\lambda_x=\lambda_y=\lambda$), 
the substitution of the ansatz (\ref{veq6})
into the Lagrangian form of the GP equation (\ref{ngp}) leads
to a quasi-particle description of the vortex center through the
Newton-type equations of motion
\begin{equation}
\ddot{x}_0 = - \frac{d\,V_{\mathrm{eff}}(x_0,y_0)}{d\,x_0}, ~~~{\rm and}~~~
\ddot{y}_0 = - \frac{d\,V_{\mathrm{eff}}(x_0,y_0)}{d\,y_0},
\label{Newton}
\end{equation}
where the effective potential is given by
\begin{equation}
V_{\mathrm{eff}}(x,y)=Q(\phi )\left[ \cos \left(\frac{4 \pi x}{\lambda}\right)
                                    +\cos \left(\frac{4 \pi y}{\lambda}\right)
\right] + 
\frac{1}{4} \Omega^{2}\left(x^2 + y^2 \right),
\label{Veff}
\end{equation}
and $Q(\phi)$ is given by a rather cumbersome expression. In our case of 
interest we will only need the following values 
\begin{equation}
\displaystyle
Q(0)=\displaystyle\frac{V_0}{4}\, \left(\frac{2b\pi^2}{\lambda^2}-1\right)\, 
\exp \left(\frac{-2b\pi^2}{\lambda^2}\right),
~~\mbox{and}~~~
\displaystyle Q\left(\frac{\pi}{2}\right)=-Q(0), 
\label{Q}
\end{equation}
Equations (\ref{Newton}) and (\ref{Veff}) indicate that the coordinates of the vortex center are prescribed by two uncoupled nonlinear oscillators (see also Ref.\ [\refcite{than}] for a similar result in the context of optics). The above, generalizes the well-known 
result\cite{dsm} that the center of a dark soliton (the 1D ``sibling'' of the vortex) behaves like a Newtonian particle in the presence of the external potential (see relevant work for dark matter-wave solitons in optical lattices in Ref.\ [\refcite{dsmol}]). Figure \ref{vortstab} depicts the vortex stability for the two detuning cases in (\ref{Q}). For $\lambda=2\pi$, $V_0=0.5>0$ and $b=1$ (by fitting ansatz to numerical solution), we have that $Q(\phi=0)=-V_0/8<0$ and $Q(\phi=\pi/2)=-Q(0)=V_0/8>0$. Therefore, a cosinusoidal OL ($\phi=0$) produces a stable vortex (see left panels of Fig.\ \ref{vortstab}) while a sinusoidal OL ($\phi=\pi/2$) induces an instability (see right panels of Fig.\ \ref{vortstab}). Note that, recently, relevant results have been obtained using different approaches\cite{nol}.

It is worth mentioning that the variational approach outlined above, although capable of capturing the stability of the vortex at the center of the trap, fails to reproduce the correct dynamics. 
More specifically, as shown in Fig.\ \ref{vortstab}, the vortex center follows an outward
spiraling motion. This spiral motion is the combination of the unstable behavior captured by the Hamiltonian dynamics of  Eqs.\ (\ref{Newton}) and (\ref{Veff}) {\em together} with the well-known precession of vortices inside the MT\cite{precession}. The precession, not captured by our approximation, predicts that, close to the trap center, a vortex rotates with an angular frequency given by\cite{precession}
\begin{equation}
\omega_{\rm prec} = \frac{-3
\omega_x\omega_y}{4 \mu} 
\ln\left(\frac{R_{\rm BEC}}{\xi}\right),
\end{equation}
where $\mu$ is the chemical potential, $R_{\rm BEC}$ is the mean transverse
dimension of the condensate and $\xi$ is the width of the vortex core, which, in fact, is the same as the healing length of the condensate\cite{review}.

\subsubsection{Discrete Models}


An interesting situation arises for strong optical lattices ($V_0 \gg \mu$) 
where the condensate is effectively transformed into a collection of 
``droplets'' (in each lattice well) that can be described by the 
spatially discrete analogue of the 
NLS\cite{ol1,tromb,konot,greiner,strong-ol}. In this case, it is possible to describe the evolution of the condensate wave function by the discrete nonlinear Schr{\"o}dinger equation (DNLS) as can be seen through a Wannier function decomposition\cite{wann1,wann2}. In non-dimensional units, the DNLS reads
\begin{equation}
i\frac{d}{dt}\phi_{\eta}+C\Delta^{(d)}_{2}\phi_{\eta}+\left\vert 
\phi_{\eta}\right\vert ^{2}\phi_{\eta}=0,  
\label{DNLS}
\end{equation}
where $\phi_{\eta}$ is the condensate wave function localized at the OL trough with coordinates $\eta=(m,n)$ and $\eta=(l,m,n)$, for the 2D and 3D cases respectively,
$C$ is the coupling constant, and $\Delta^{(d)}_{2}$ stands for the
discrete Laplacian in $d$ dimensions: 
$\Delta^{(2)}_{2}\phi _{m,n}=\phi _{m+1,n}+\phi _{m,n+1}+\phi
_{m,n-1}+\phi _{m-1,n}-4\phi _{m,n}$ and 
$\Delta^{(3)}_{2}\phi _{l,m,n}=\phi _{l+1,m,n}+\phi _{l,m+1,n}+\phi
_{l,m,n+1}+\phi _{l-1,m,n}+\phi _{l,m-1,n-1}+\phi _{l,m,n-1}-6\phi _{l,m,n}$.
In Refs.\ [\refcite{nova,vor3d,vor2d}] stationary solutions of (\ref{DNLS}) are
sought by considering $\phi_{\eta}=\exp (-i\mu_{0} t)\,u_{\eta}$, where $\mu_{0}$ is the dimensionless chemical potential of the condensate, that yields to
the time-independent equation:
\begin{equation}
-\mu_{0} u_{\eta}=C\Delta^{(d)}_{2}u_{\eta}+\left\vert u_{\eta}\right\vert
^{2}u_{\eta},  \label{standing}
\end{equation}
where $|u_{\eta}|^2$ is proportional to the atomic density at the $\eta$-th
trough. Since Eq,\ (\ref{standing}) has a scale invariance,
$\mu$ can be fixed arbitrarily. 
As in Refs.\ [\refcite{nova,vor3d,vor2d}], 
by using an appropriate initial discrete vortex ansatz, it is possible to
find numerical solutions to Eq.\ (\ref{standing}). Then, by applying continuation-type
methods, together with linear stability computations, the branches of 
existence and stability for discrete vortices of different charges in two- ($d=2$) and
three-dimensional ($d=3$) settings can be obtained.

\begin{figure}[th]
\centerline{
(a)~~~~~~~~~~~~~~~~~~~~~~~
(b)~~~~~~~~~~~~~~~~~~~~~
(c)~~~~~~~~~~~~~~~~~~~~~~
(d)~~~~~~~~~~~~~~~~~~
\\[-2.0ex]
}
\centerline{
\epsfig{figure=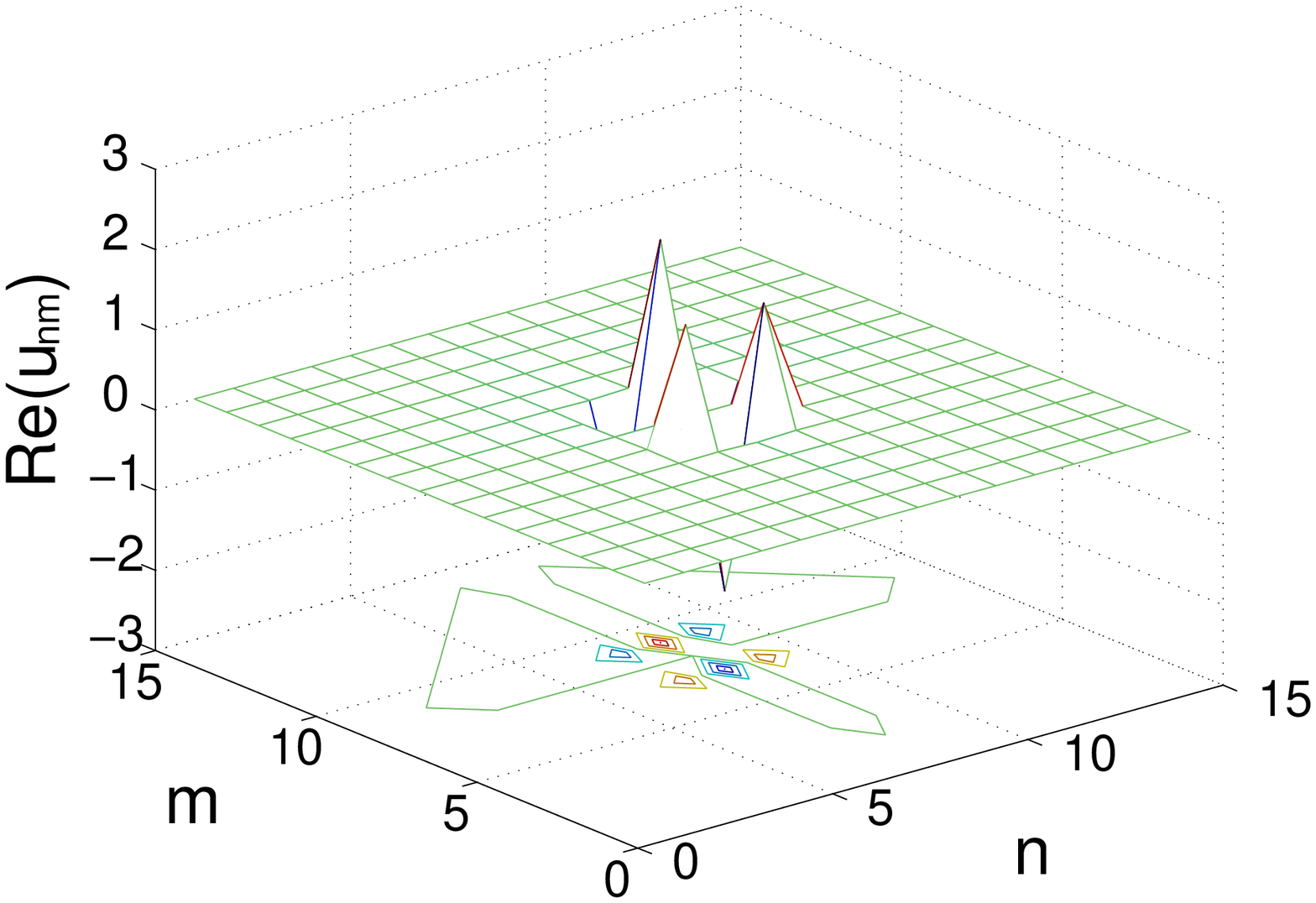,width=3.0cm,angle=0,silent=}
\epsfig{figure=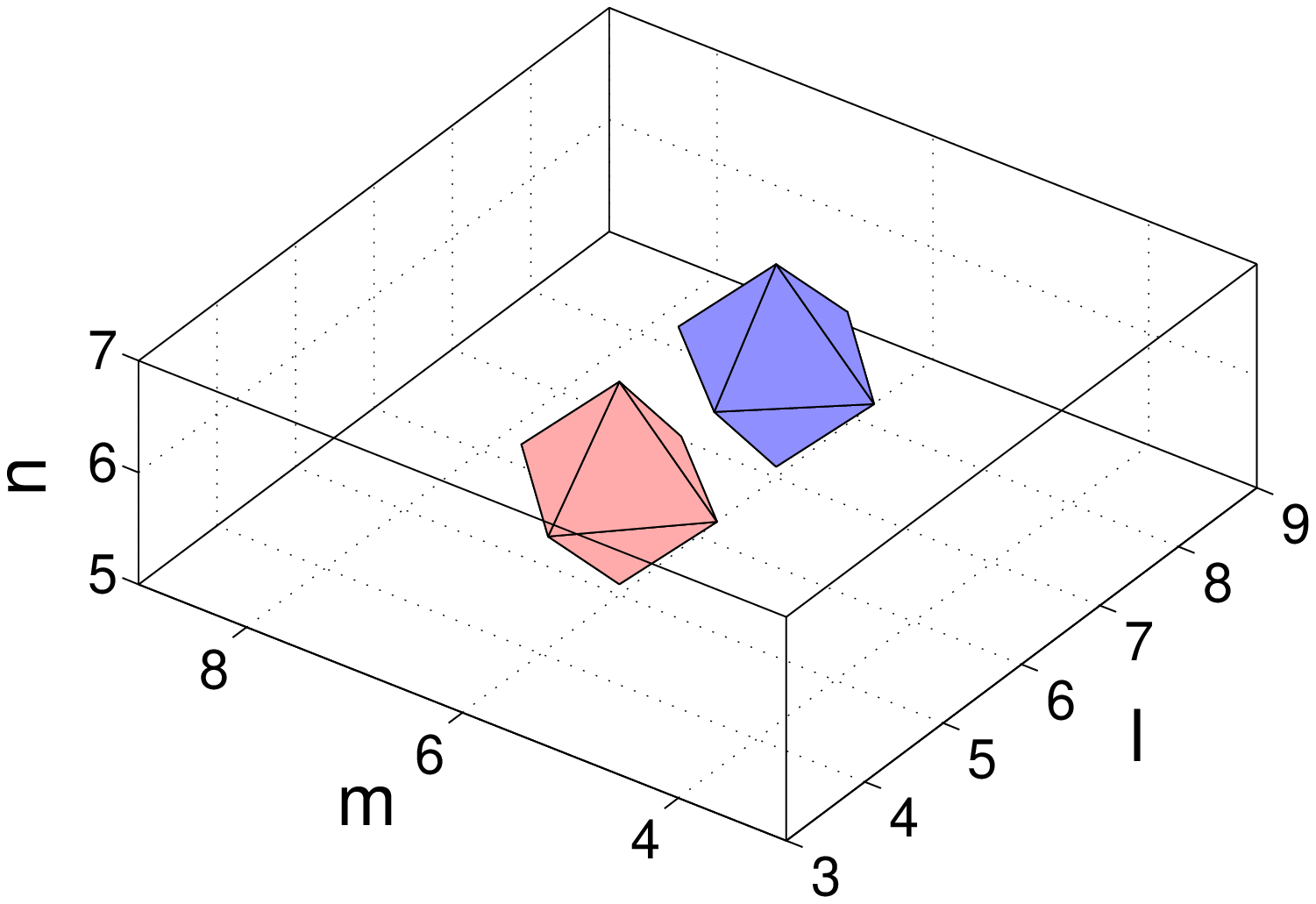,width =3.0cm,angle=0,silent=}
\epsfig{figure=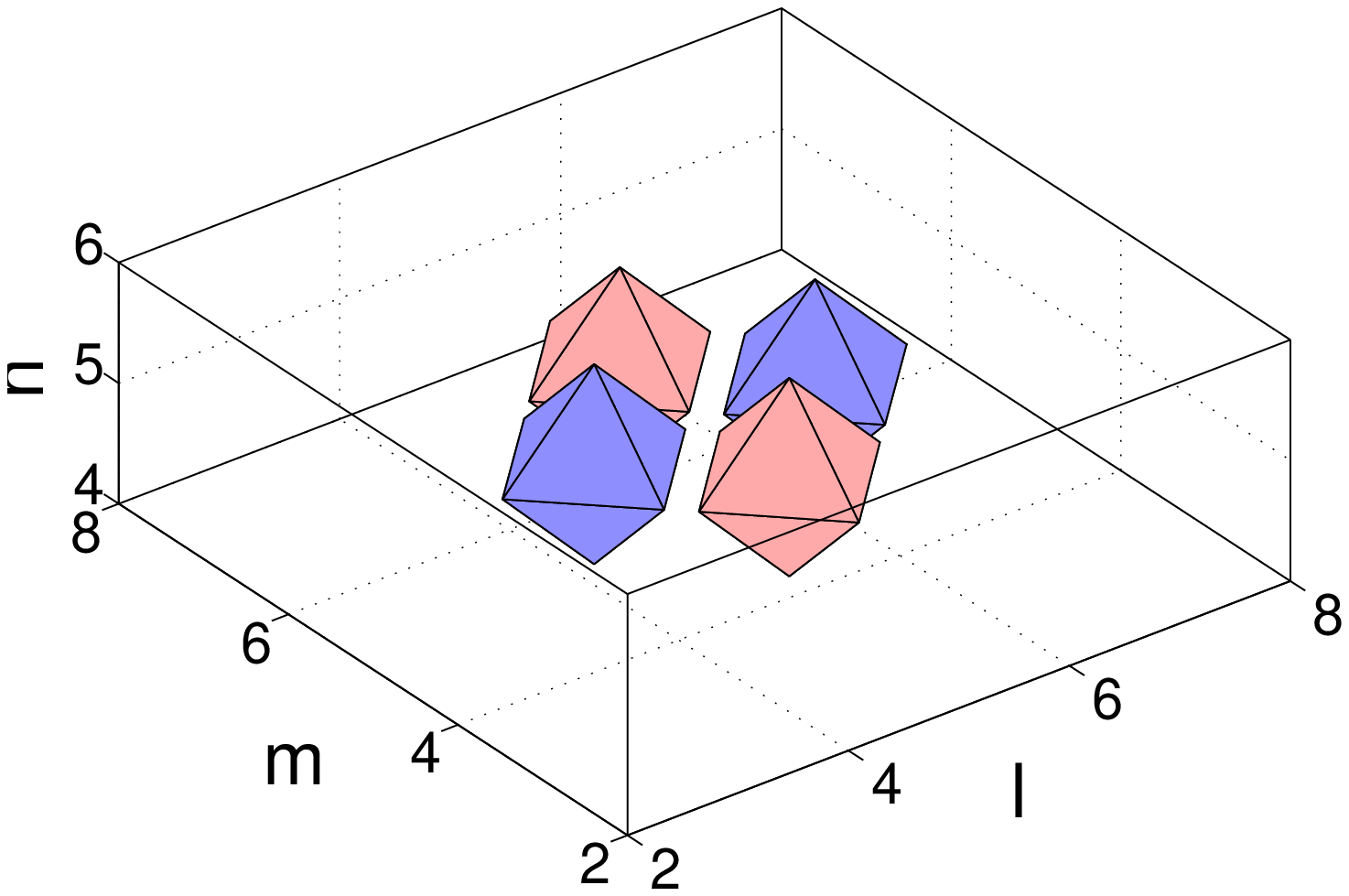,width =3.0cm,angle=0,silent=}
\epsfig{figure=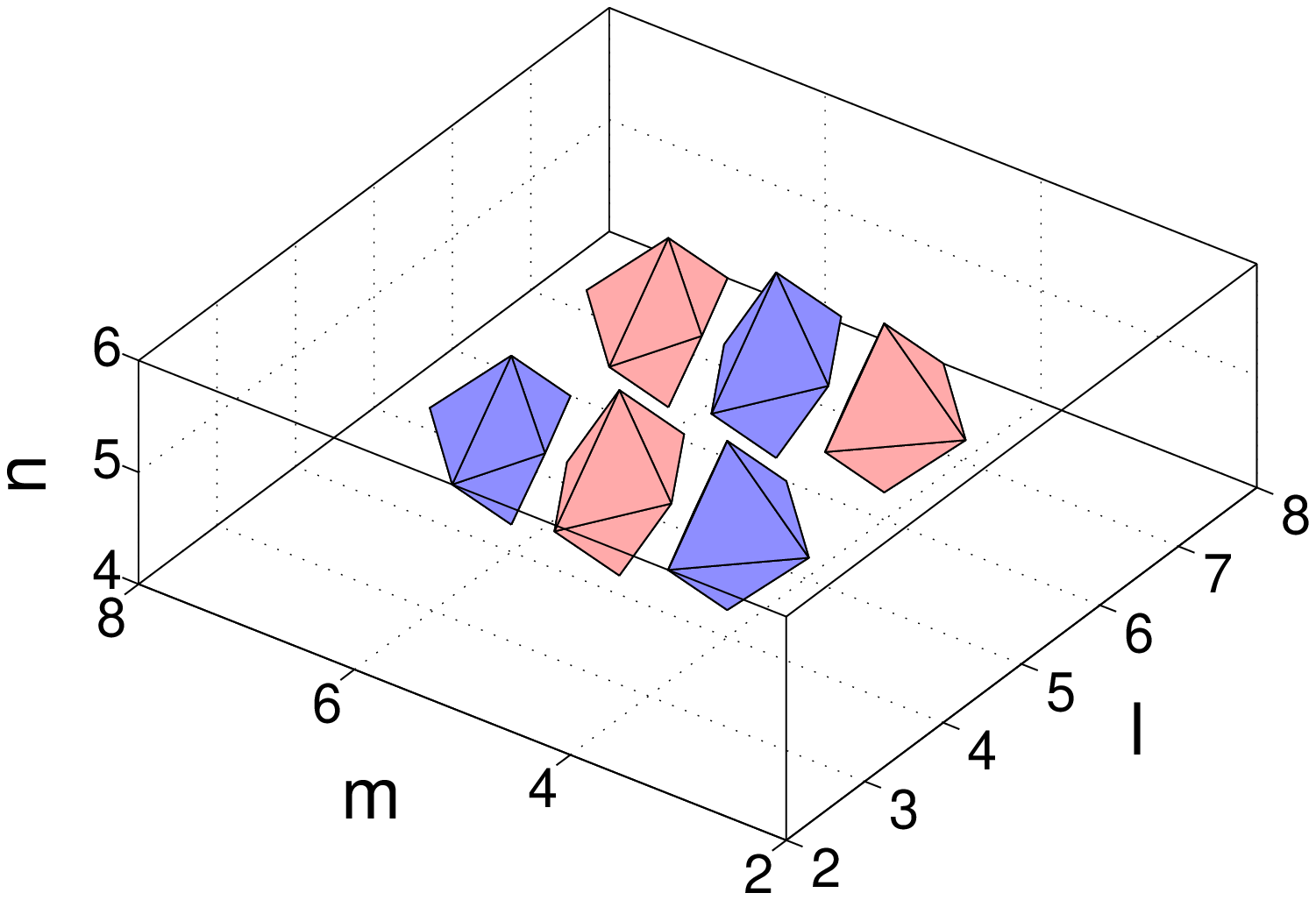,width =3.0cm,angle=0,silent=}
}
\centerline{
\epsfig{figure=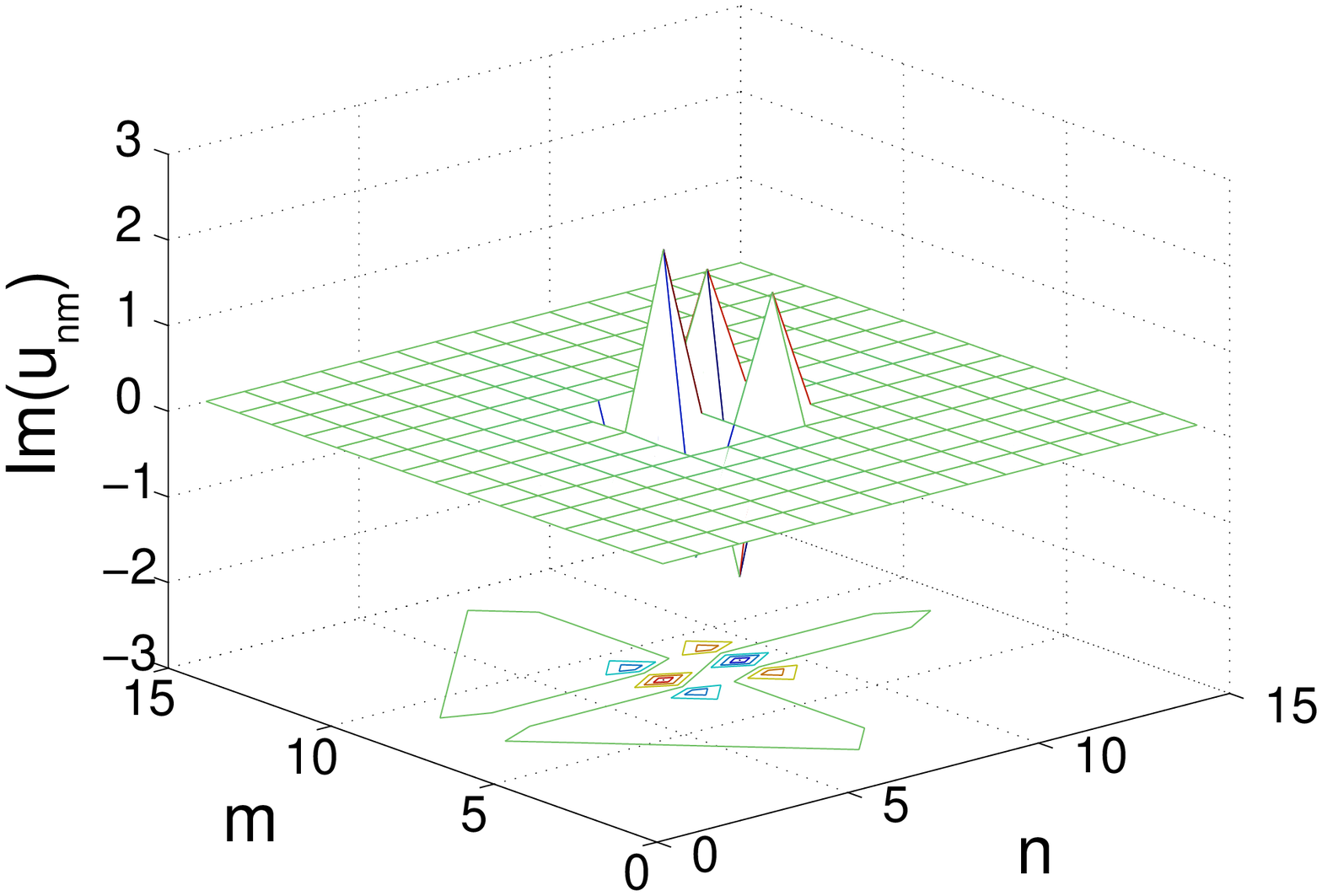,width=3.0cm,angle=0,silent=}
\epsfig{figure=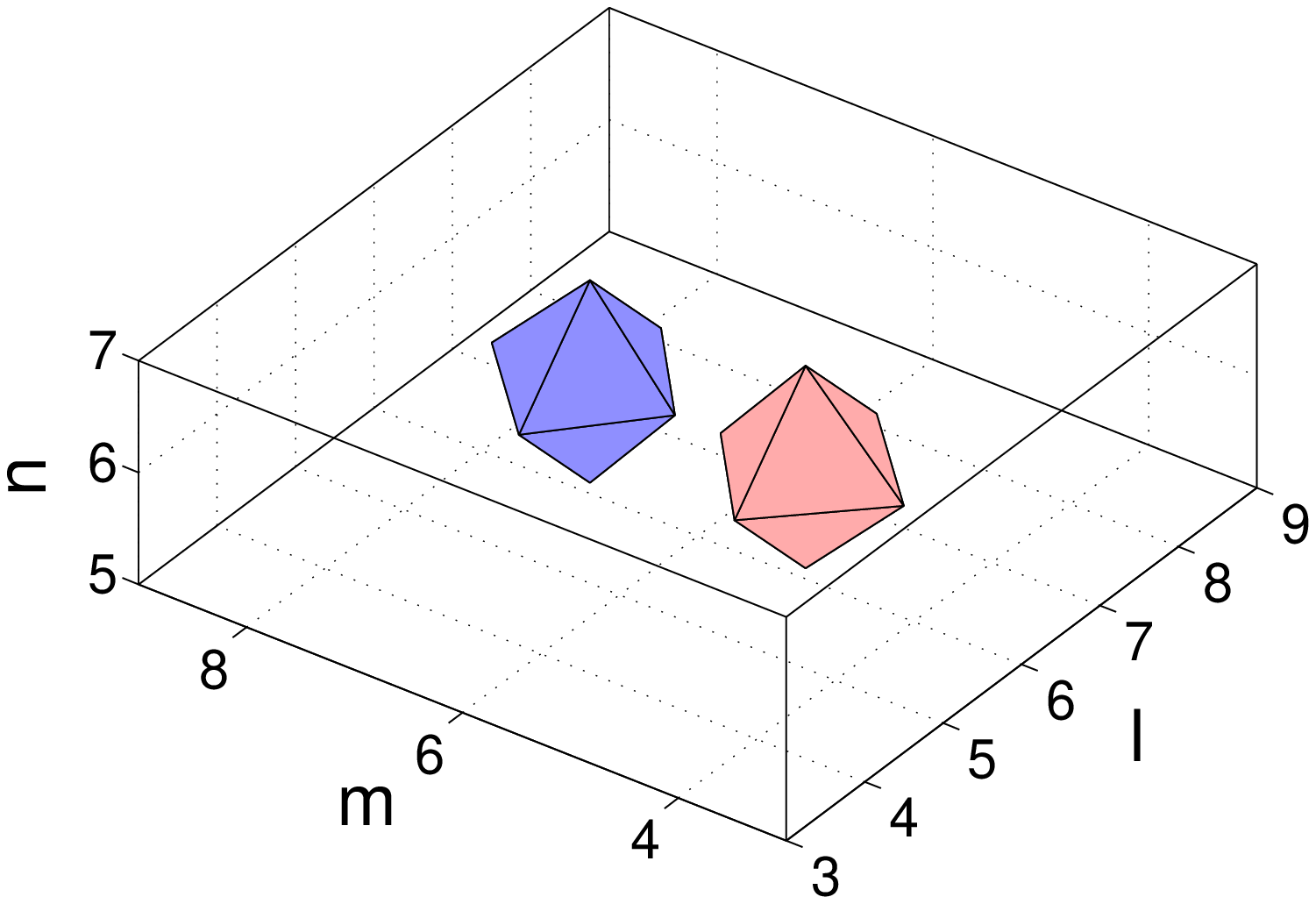,width =3.0cm,angle=0,silent=}
\epsfig{figure=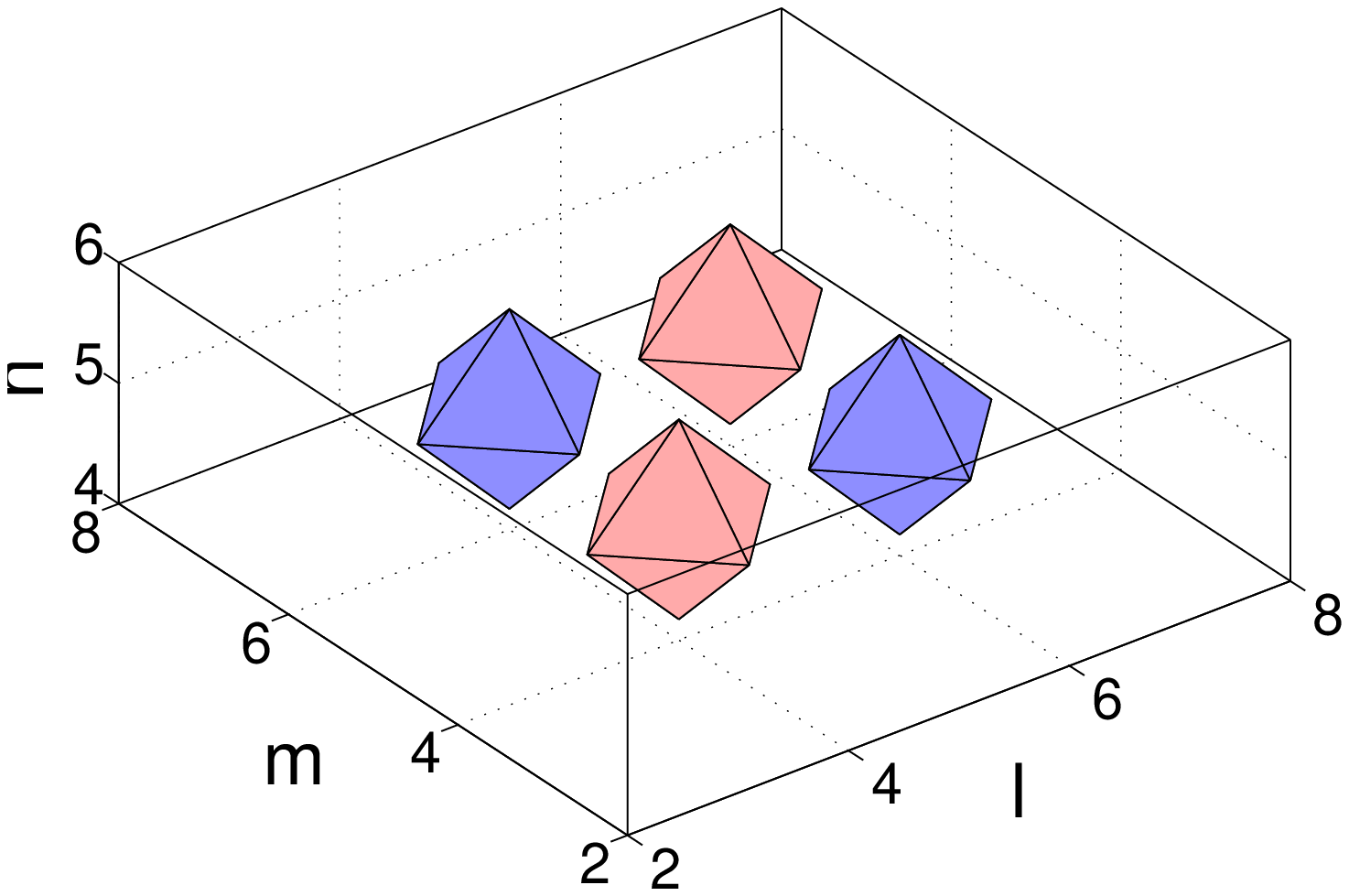,width =3.0cm,angle=0,silent=}
\epsfig{figure=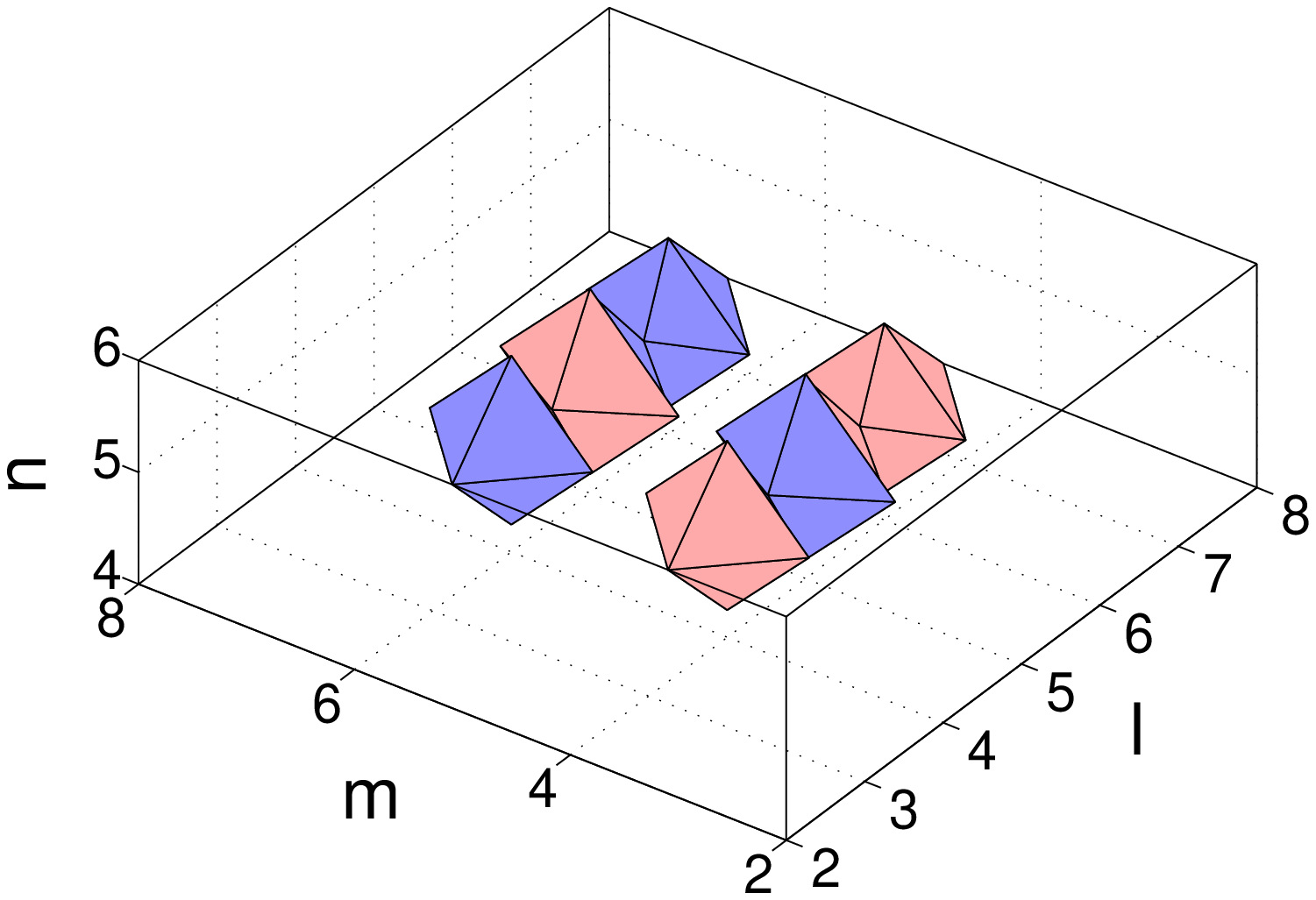,width =3.0cm,angle=0,silent=}
}
\caption{
Vortices in the discrete nonlinear Schr{\"o}dinger equation
in two ($d=2$) and three ($d=3$) dimensions.
The top (bottom) row depicts the real (imaginary) part
of the stationary solution (cf.\ Eq.\ (\ref{standing}))
with vorticity $S$.
For $d=3$ (b,c,d) the plots depict the contour plots 
for ${\rm Re}(u_{l,m,n})=\pm$constant (top row) and
${\rm Im}(u_{l,m,n})=\pm$constant (bottom row).
The parameters are, from left to right: 
(a) $d=2$, $S=3$, $C=0.02$ 
(b) $d=3$, $S=1$, $C=0.7$, constant=0.5
(c) $d=3$, $S=2$, $C=0.01$, constant=0.25  and
(d) $d=3$, $S=3$, $C=0.01$, constant=0.25. 
}
\label{discretevort}
\end{figure}

The stability results presented in Refs.\ [\refcite{nova,vor2d}], for vortices
of charge $S$ (e.g., for dimensionless chemical potential $\mu_{0}=-4$), 
can be summarized in the following stability table:
\begin{center}
\begin{tabular}{|c|c|c|c|c|}
\hline
$d$  & $S=0$ & $S=1$ & $S=3$ & $S=2$ \\[0.0ex]
\hline
2 & $C\lesssim C_{\rm cr}^{(2,0)}\approx 4 $ & $C\lesssim C_{\rm cr}^{(2,1)}\approx 1.6$ 
  & $C\lesssim C_{\rm cr}^{(2,3)}\approx 0.398$ & U\\[0.0ex]
\hline
\end{tabular}
\end{center}
where the approximate region for stability is indicated (the letter ``U'' denotes 
unstable for all values of $C$). As indicated,
no stable 
vortices with $S=2$  were obtained\cite{pgk4-boris}. Nonetheless, 
Eq.\ (\ref{standing}) admits {\em real} stationary solutions
---generated by the real part of the corresponding 
genuine vortex (which is complex)---
that are stable for sufficiently weak coupling. These real solutions, 
the so-called quasi-vortices, correspond to 
quadrupoles ($S=2$) and octupoles ($S=4$)\cite{nova,vor2d}. 
Similar results were found also for the three-dimensional case in Ref.\ [\refcite{vor3d}].
In Fig.\ \ref{discretevort} we depict a few examples of discrete vortices for different vorticities in two- and three-dimensions.

It is important to mention that the discreteness is responsible for inducing 
the stability of vortices in three-dimensional settings that otherwise 
(in the continuum model) are strongly unstable. A natural question that 
arises is the fate of unstable solutions. For example, we have observed that,
in two dimensions, a vortex with $S=3$ and $C=0.618> C_{\rm cr}^{(2,3)}$
may decay into a configuration consisting of a combination of a stable 
soliton ($S=0$) and a stable $S=1$ vortex (here $C<C_{\rm cr}^{(2,0)}$ and $C<C_{\rm cr}^{(2,1)}$). On the other hand, we have observed a striking effect where, in three-dimensions, an unstable $S=2$ vortex decays into a {\em stable} $S=3$
vortex (for $C=0.01<C_{\rm cr}^{(3,3)}$), thereby increasing the
total topological charge instead of decaying to
a combination of lower order ($S=0,1$) vortices. It should be noted that 
this change of vorticity is possible in the discrete lattice model, in which the angular momentum is not a dynamical invariant.
Another noteworthy vortex solution also found in Refs.\ [\refcite{vor3d}]
consists of a complex of two mutually orthogonal $S=1$ vortices
(one in each component) in the discrete version of a two-component condensate 
(cf.\ Section \ref{sec:twocomp}).

\section{Two Vortices}\label{sec:TwoVortices}

\subsection{One-Component BEC: Interactions}

\begin{figure}[th]
\centerline{
~
\epsfig{figure=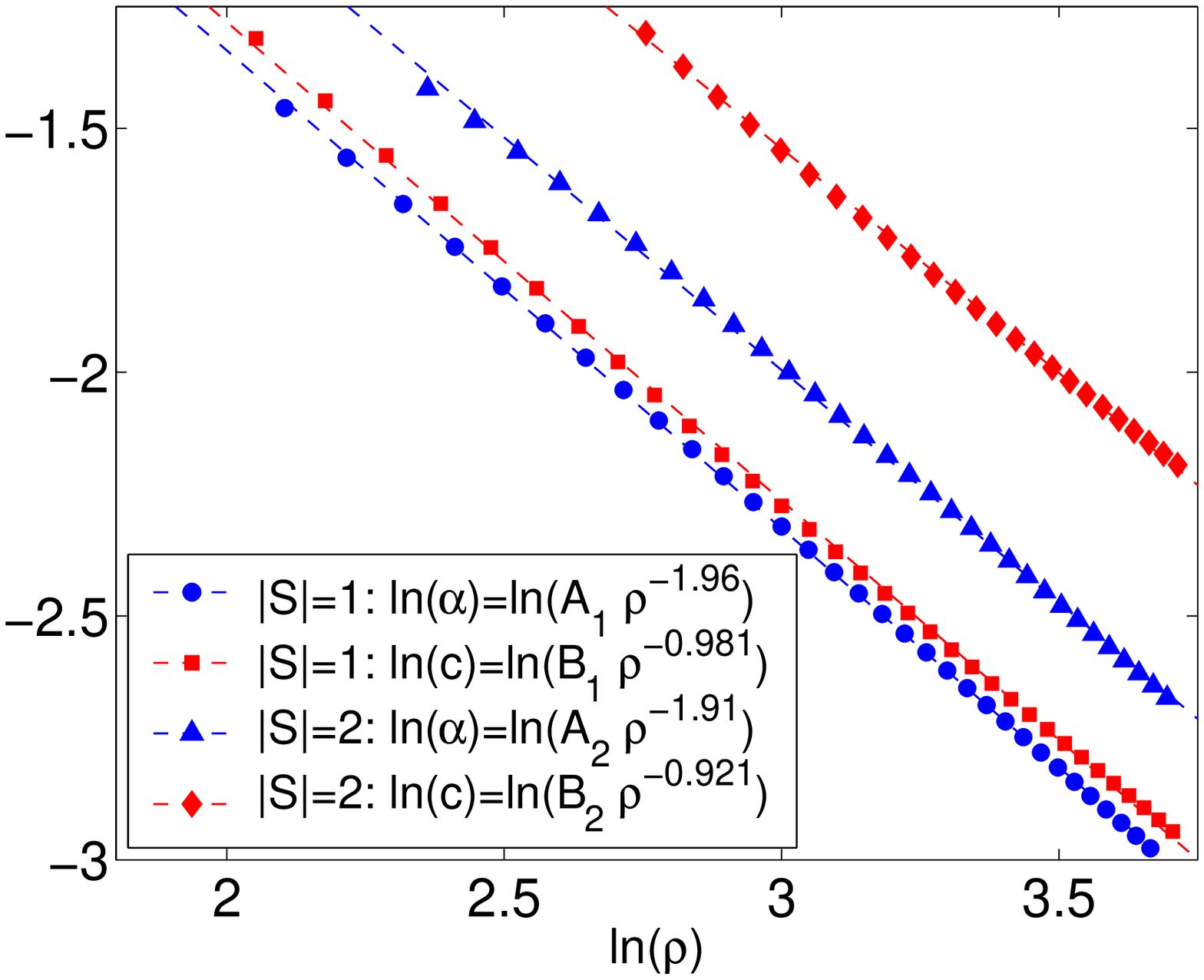,width=4.0cm,height=3.5cm,angle=0,silent=}
~
\epsfig{figure=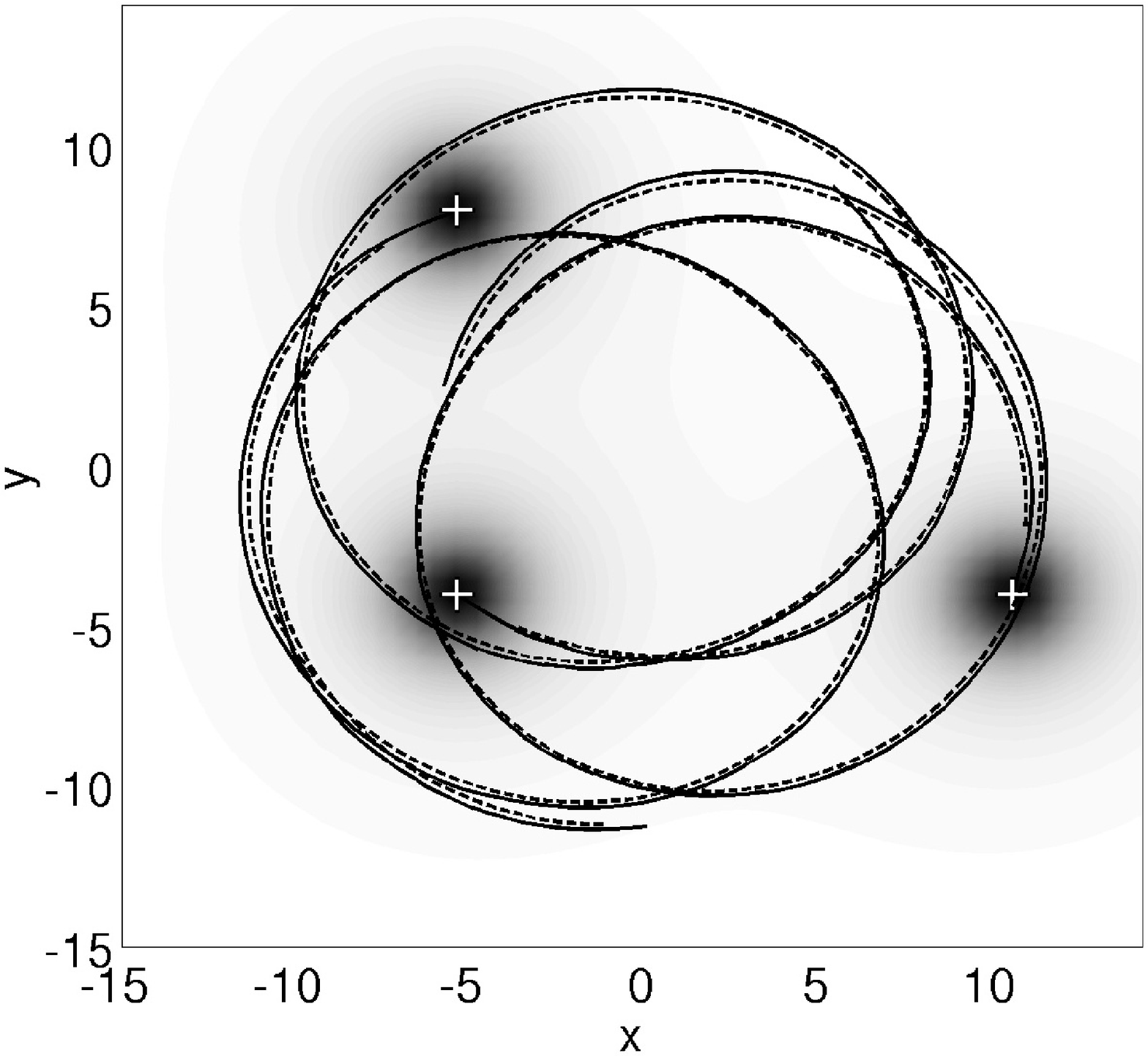,width=3.7cm,height=3.7cm,angle=0,silent=}
\epsfig{figure=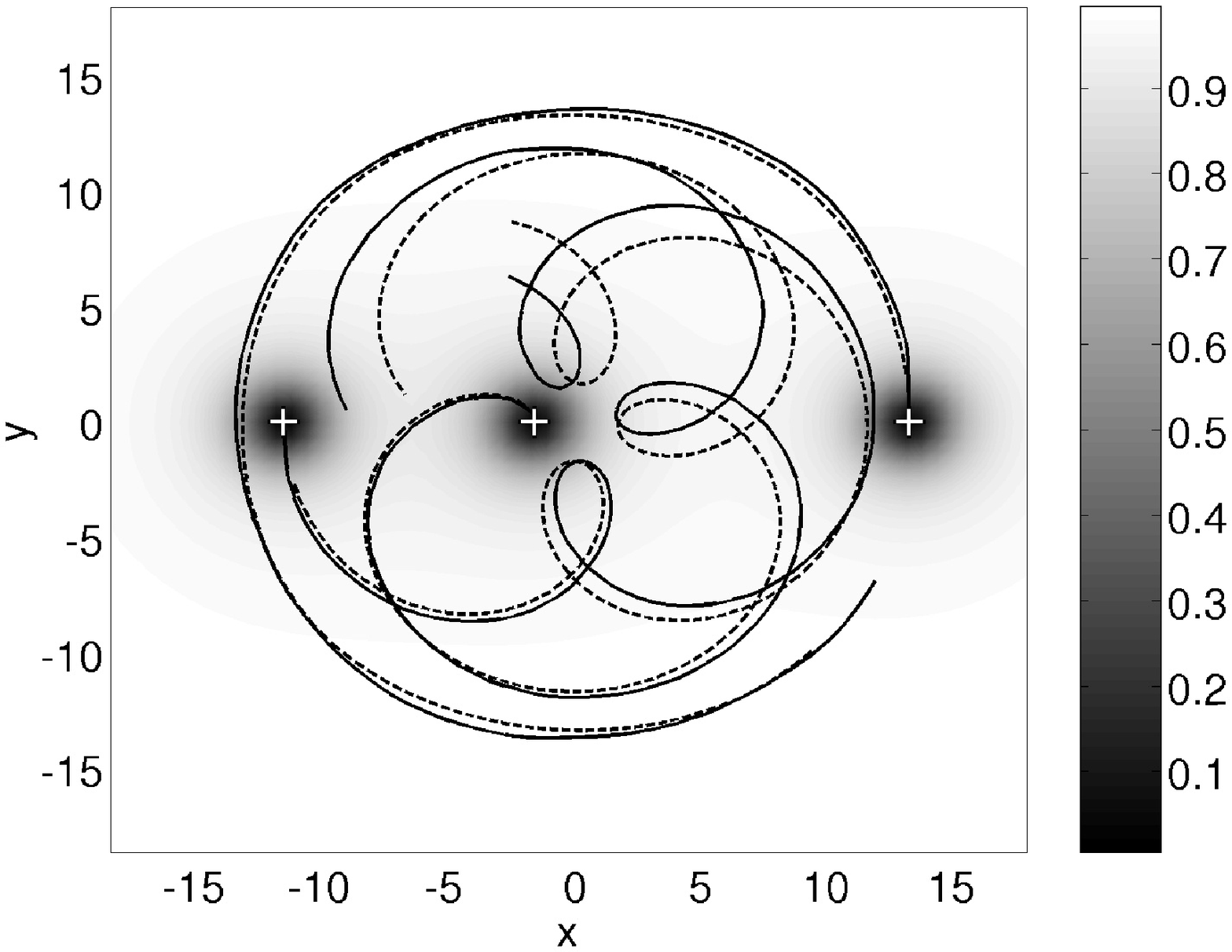,width=4.7cm,height=3.9cm,angle=0,silent=}
}
\caption{Left: linear $c(\rho)$ (angular $\alpha(\rho)$) velocity for two 
vortices of opposite (equal) charge $|S|=|S_1|=|S_2|=1$ and $|S|=|S_1|=|S_2|=2$ 
as a function of their separation distance $\rho$.
Middle and right panels depict two configurations of three interacting
vortices of charge $S=1$. The solid and dashed line represent the trajectories 
of their centers from direct numerical integration of (\ref{peq1}) and 
from the quasi-particle approximation of (\ref{odes}), respectively.
}
\label{vortode}
\end{figure}

Up to this point we have dealt with the generation and stability of vortices. Let us now focus on the important issue of vortex-vortex interactions. As is the case also for fluid vortices, two BEC vortices of opposite charge travel parallel to each other at constant speed $c$, while  
vortices of same charge rotate at constant angular speed 
$\alpha$ \cite{Fetter:65,Fetter:66}. 
From direct numerical integration of Eq.\ (\ref{ngp}) 
we have been able to characterize the dynamics
of interacting vortices
in the absence of any trapping potential ($\Omega=0$).
With an appropriate initial phase mask (see Section \ref{subsec:Generation})
we seed two vortices, with respective vorticities $S_1$ and $S_2$, at a 
desired distance from each other.
Then, we fitted a Pad\'e approximation\cite{pade} to find the vortex centers
during evolution, and numerically
extracted the linear (angular) velocities for vortices
of opposite (equal) charge as a function of their separation distance $\rho$
(see left panel in Fig.\ \ref{vortode}).
We performed our experiments for same and opposite charge vortices
with vorticities $|S_1|=|S_2|=1$ and $|S_1|=|S_2|=2$.
The angular and linear velocities of interacting vortices seem
to nicely obey power laws for a wide range of separations ($10<\rho<50$):
$c(\rho) \sim \rho^{-1}$ and $\alpha(\rho) \sim \rho^{-2}$ (see left panel
of Fig.\ \ref{vortode}).

The next step to describe the dynamics of interacting vortices
consists on identifying an appropriate Lagrangian that gives rise
to the correct dynamics. In order to achieve this, it is crucial to note
that BEC vortices are known to obey 
first order, kinematic equations\cite{fetter}, contrary
to what is known for the Newtonian (second order) equations describing
other types of solitary wave center of mass dynamics\cite{Progress}. 
%
To circumvent this obstacle, we need to construct an interaction Lagrangian
that gives the ``correct'' vortex dynamics through its Euler-Lagrange
equations. 
Such a Lagrangian for two vortices, with {\em same} vorticity $S=S_m=S_n$,
may be given in the form
\begin{equation}
\label{LagrangianMD2}
L_{m,n}
\sim
%
%
{\rm det}(\dot{\vec{\,r}}_n,\vec{\,r}_m)+
{\rm det}(\dot{\vec{\,r}}_m,\vec{\,r}_n)
- A\, S \ln(\rho)
\end{equation}
where $\vec{r}_m=(x_m,y_m)^T$ is the position vector of vortex $m$
and $A$ is a constant.
%
%
%
For vortices of opposite charge one needs to multiply the velocities
for each vortex by their respective charge. This is equivalent to 
including the vortex charge in the definition of the Poisson brackets
for the vortex
interaction Hamiltonian (cf.,\ Ref.\ \refcite{Borisov:98} and references
therein).

Let us explicitly write the equations of motion for same charge
vortices. In this case, the pairwise Lagrangian 
(\ref{LagrangianMD2}) gives rise to the
well known equations of motion for two (fluid) point vortices centered at 
$\vec{r}_n=(x_n,y_n)^T$ and $\vec{r}_m=(x_m,y_m)^T$:
\begin{equation}
\label{odes}
\begin{array}{rcl}
\ds \dot x_m &=& \ds -AS\,\, \frac{y_m-y_n}{2\rho^2} \\[2.0ex]
\ds \dot y_m &=& \ds +AS\,\, \frac{x_m-x_n}{2\rho^2},
\end{array}
\end{equation}
where $\rho=\sqrt{(x_m-x_n)^2+(y_m-y_n)^2}$ is inter-vortex distance 
and $A=A_1\approx 3.96$ for $S=1$ and $A=A_2\approx 7.80$ for $S=2$ 
is a constant determined from the numerics 
(cf.\ left panel in Fig.\ \ref{vortode}).
As expected we have $2A_1=7.92\approx A_2$, namely, the effects of
an $S=2$ vortex are equivalent to the superposition
of the effects of two nearby $S=1$ vortices.
For opposite charge vortices traveling parallel to each other at constant speed
$c$, our numerics predict that $c(\rho)=B \rho^{-1}$ with $B=B_1\approx 2.15$
and $B=B_2\approx 4.43$ for singly- ($|S|=1$) and doubly-charged ($|S|=2$)
vortices respectively. As for same charge vortices,
the relationship $2B_1=4.30\approx B_2$ holds approximately.

Approximation (\ref{odes}) treats vortex pairs as quasi-particles with 
interacting potentials. It is important to mention that this approximation
to the dynamics deteriorates when the vortices get too close to each other.
Nonetheless, we have checked that the approximation remains valid down to
separation distances of the order of the healing length (i.e., the 
width of the vortices) for $|S|=1$ vortices. For $|S|=2$,
the two vortices tend to split into two
pairs of $S=1$ vortices\cite{Mottonen:04} when the separation distance 
was decreased below approximatively $\rho=10$.

In the middle and right panel of Fig.\ \ref{vortode} we present
two examples with three interacting vortices.
The solid line represents the 
actual orbits obtained from direct numerical integration of Eq.\ (\ref{peq1}),
while the dashed line depicts the orbits using the quasi-particle approximation
(\ref{odes}). As it can be observed from the figure, the superposition of 
all the interactions, given by (\ref{odes}), for the three vortices is
in good agreement with the full model. It is worth mentioning that the
case depicted in the right panel of the figure corresponds to an orbit
with an initial condition close to an unstable configuration. This explains
the larger discrepancy for this case when compared to the middle panel.

We note in passing that a more general approach towards computing 
vortex interactions may involve the use of a Lagrange multiplier at the level
of the PDE (initialized with two vortices) which maintains the
distance between the vortices fixed. Then, the force associated with
the relevant multiplier is what is needed to balance the interaction
force between the vortices and hence can also be used to infer the 
interaction potential.

\subsection{Two-Component BECs: Dipole Bound States}\label{sec:twocomp}

A very relevant generalization of the class of physical systems that we have discussed so far,
and of the solitary waves they can support, concerns the case of coupled multi-component BECs. There has recently been a considerable volume of work relevant to the properties of coupled BECs ranging from the study of ground state solutions\cite{shenoy,esry} to 
small-amplitude excitations\cite{excit}. Furthermore, the formation of various structures such as domain-walls\cite{Marek,haelt,cross,obsantos}, 
bound dark-dark\cite{obsantos}, 
dark-bright\cite{anglin}, 
dark-antidark, dark-gray, bright-antidark and bright-gray soliton complexes\cite{epjd}, 
as well as spatially periodic states\cite{decon} was also predicted. On the other hand, 
experimental results have been reported for mixtures of different spin states of $^{87}$Rb 
(see Ref.\ [\refcite{myatt}]) and mixed condensates\cite{dsh,stamp}. It is relevant to also mention the efforts towards the realization of two-component BECs from different atomic species, such as $^{41}$K--$^{87}$Rb (see Ref.\ [\refcite{KRb}]) and 
$^{7}$Li--$^{133}$Cs (see Ref.\ [\refcite{LiCs}]). 

Typically, the generalized mean field model for two coupled BECs involves two nonlinearly coupled GP equations. However, in experiments with a radio-frequency (or an electric field) coupling two separate hyperfine states\cite{myatt}, the relevant model also involves a linear coupling between the wavefunctions. The governing normalized equations are then of the form:
\begin{eqnarray}
i \psi_{1t}=\left[-\frac{1}{2} \Delta+V+ g_{11}|\psi_{1}|^{2}+g_{12}|\psi _{2}|^{2}\right]\psi_{1}+ \alpha \psi_{2},\label{lceqa} \\
i \psi_{2t}=\left[-\frac{1}{2} \Delta+V+ g_{12}|\psi _{1}|^{2}+
g_{22}|\psi _{2}|^{2}\right]\psi_{2}+ \alpha \psi_{1},
\label{lceqb}
\end{eqnarray}
where the intra- and 
inter-species interactions are characterized by the coefficients 
$g_{jj}$ ($j=1,2$) and $g_{12}$ respectively, 
while  $\alpha$ denotes the strength of the 
radio-frequency (or electric field) coupling.
In the recent work of Ref.\ [\refcite{bernard}], the special case of
$g_{11}=g_{22} = g_{12} \equiv g$ was examined. The latter
can be written in a vector form as:
\beq\la{eqn:stark}
i {\bpsi}_t -\alpha P \bpsi =
-\frac{1}{2}\Delta {\bpsi}+({\bpsi}^\dagger G \bpsi)\bpsi+V({\bf x})\bpsi,
\eeq
where
\beq
P=\left(
\ba{cc}
0&1\\
1&0
\ea
\right)
\eeq
and $G=g {\bf I}$, with ${\bf I}$ being the unit matrix.
In that case, as was also previously known in optics\cite{potasek},
one can make a unitary transformation:
\beq\la{unit}
\bpsi=U(t)\bpsi_0=e^{-i \alpha P t}\bpsi_0=\left(
\ba{cc}
\cos(\alpha t)& -i \sin(\alpha t)\\
-i \sin(\alpha t) & \cos(\alpha t)
\ea
\right)\bpsi_0. 
\eeq
Then the original set of equations is transformed into:
\beq\la{eqn:nls}
i\bpsi_{0t}=-\frac{1}{2} \Delta \bpsi_{0}+(\bpsi_0^\dagger G \bpsi_0)\bpsi_0+
V({\bf x})\bpsi_0,
\eeq
i.e., the linear coupling can be completely eliminated from the 
equations.
Notice that this special case of approximately equal scattering
length coefficients is relevant to the experiments performed
with two spin states of $^{87}$Rb (see Ref.\ [\refcite{dsh}]), where 
$g_{11}:g_{12}:g_{22}=1.03:1:0.97$.

\begin{figure}[tbp]
\begin{center}
\hskip-0.1cm
\epsfig{file=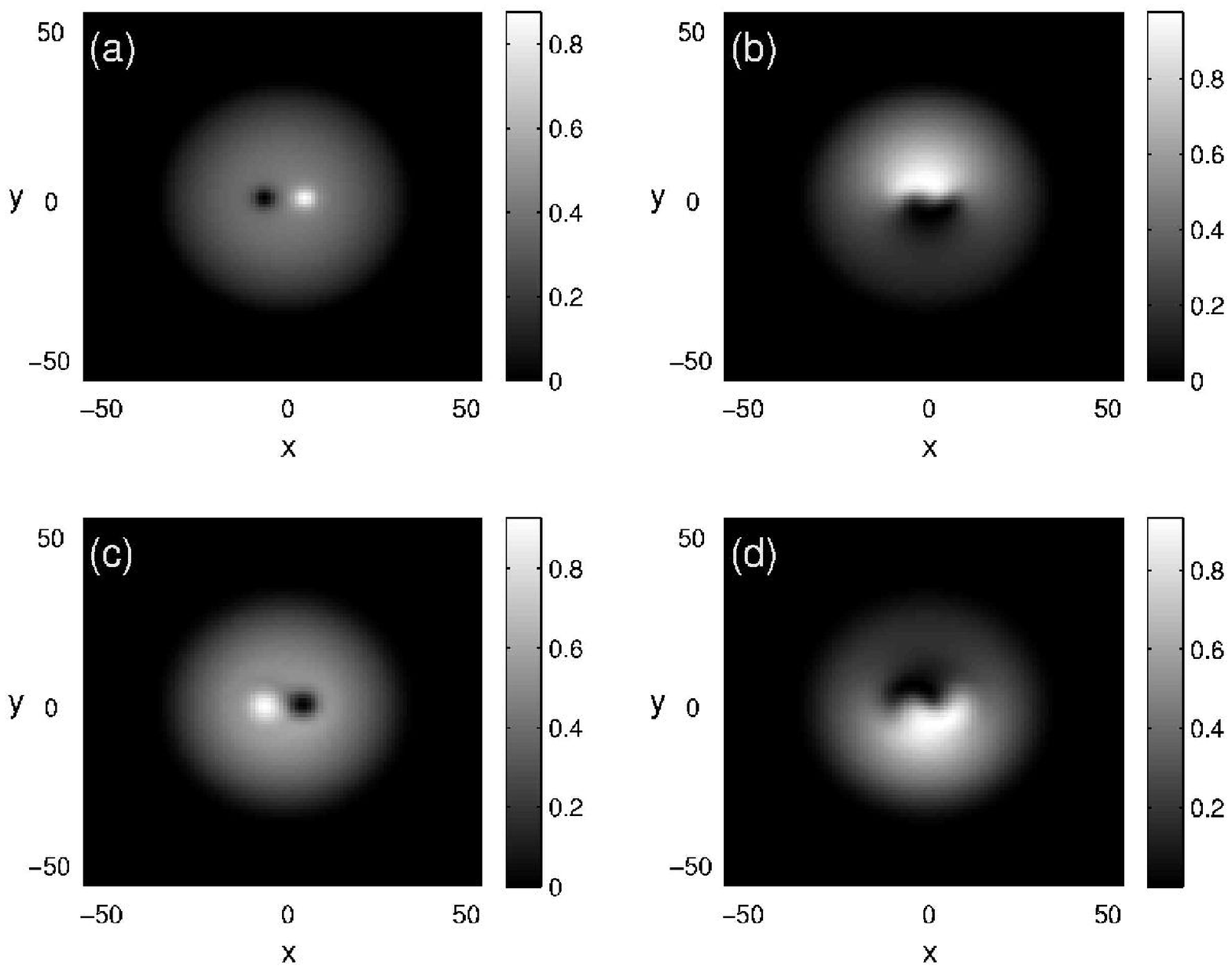, width=2.4in, height=5.2cm ,angle=0,silent=}
~~~
\epsfig{file=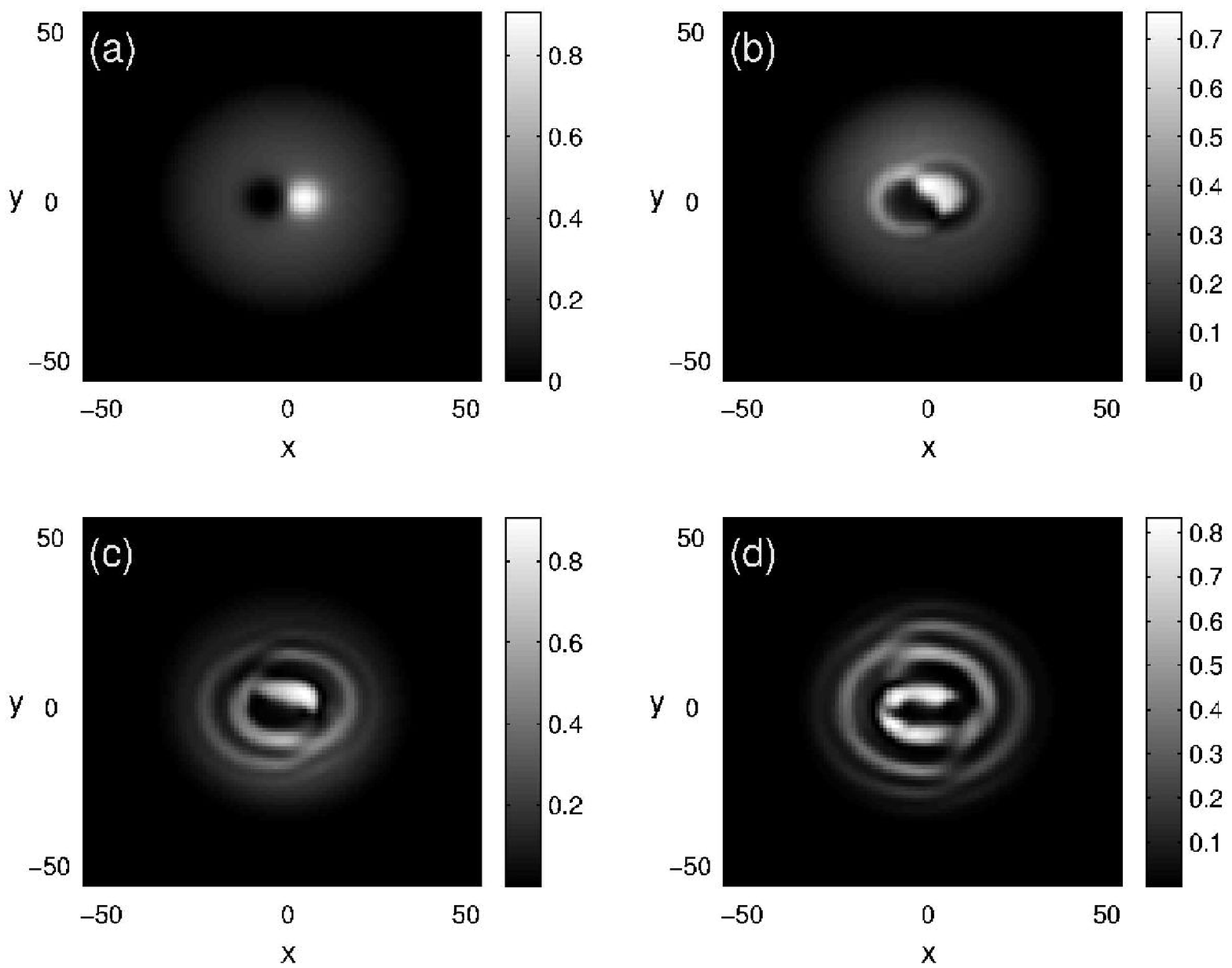, width=2.4in, height=5.2cm ,angle=0,silent=}
\end{center}
\caption{
Left panels: Contour plots of $|\psi_{1}|^{2}$ for two coupled 
vortices, initially placed at $x= \pm 5$, for $t=0$ (a), $T/4$ (b), 
$T/2$ (c), and $3T/4$ (d), 
with $T=\pi/ \alpha \approx 15.7$ ($\alpha=0.2$);
$\Omega=0.045$, $\Delta \equiv g_{11} g_{22}-g_{12}^2=-9 \times 10^{-4}$ ($^{87}$Rb). The vortices ``interchange locations'' (in a structure resembling a spiral wave).
Right panels: Same as the left but for $\Delta=-3$ ($g=1$, $h=2$). The configuration breaks up forming spiral patterns.}
\label{mfig4}
\end{figure}

In the system of nonlinearly coupled GP equations for $\bpsi_0$,
one can construct a dipole configuration with a pair of vortex
structures, see Fig.\ \ref{mfig4}. This configuration was obtained
in the figure, by means of imaginary time integration in the
absence of linear coupling, starting with one component having a
vortex centered at $(5,0)$, while the other has a vortex at $(-5,0)$.
After
the configuration relaxes to the stationary vortex pair solution of 
Eqs.\ (\ref{lceqa})--(\ref{lceqb}), one can turn on the linear coupling and 
obtain a spiral rotation between the vortices resembling a spiral wave. 
While this solution is {\it exact} for $h \equiv
g_{12}/g_{11}=1$ (assuming $g_{11}=g_{22}$), it also persists for
non-unit values of $h$. In particular, the spiral structure persists
for 
$h < h_{c}=1.32$ beyond which the regularity 
of the Rabi oscillations of matter between the components
is destroyed. In this case, the breakup leads to the 
formation of spiral patterns in the condensate\cite{bernard}.

\section{Vortex Lattices}\label{sec:VortexLattices}

We now turn to a ``thermodynamic limit'' level which is also, however,
particularly relevant to BEC experiments.  Rapid rotation of 
2D BECs induces the generation of many 
vortices\cite{fetter,newton,rokshar} that typically settle into
ordered lattices\cite{vort2,vort3,latt1,latt2,latt3,engel1,engel2}.
Particularly enticing in this respect are the available pictures
of such lattices and their (practically perfect)
triangular patterns\cite{web:VL}. These are the so-called
Abrikosov lattices\cite{abrikosov}, that were long ago predicted
in the theory of superconductivity (and that are cited as the 
prediction that earned their discoverer the Nobel prize in Physics
in 2003). In the context of
type-II superconductors, free energy arguments 
can be used to demonstrate that the triangular lattice is the most
energetically favorable (ground-state) configuration\cite{tkach}.

\begin{figure}[ht]
\centerline{
\epsfig{file=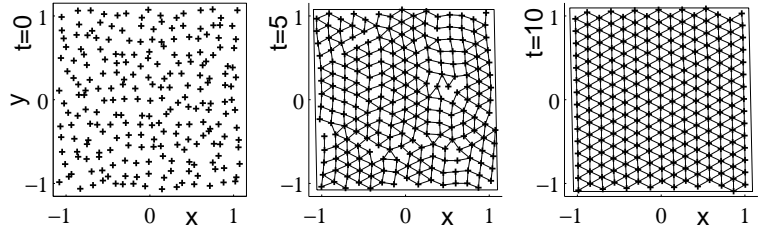, width=4in ,angle=0,silent=}
}
\caption{Typical vortex lattice MD simulation with a random initial pattern
of $N_v=225$ same-charge vortices.
After a short transient ($t<8$), the configuration equilibrates into a
triangular lattice ($t=10$).
\label{pfig8}}
\end{figure}

It is quite interesting 
in this context, to study 
vortex lattices in the framework of Eq.\ (\ref{ngp}) 
both in the presence of the external magnetic trap (MT),  
as well as in the one of the optical lattice (OL). 
Naturally, it is relevant to 
perform direct simulations of such 
lattices for different (external potential) parameter values.
However, a quite 
relevant alternative tool in order 
to study the ground states of such configurations
and their structural transitions is the use of molecular dynamics (MD)
techniques such as the Parrinello-Rahman (PR) method\cite{Rahman-Parrinello:a}. 
In the PR dynamics, not only are the positions
of the vortices (viz.\ particles) evolved in time, but so are the coordinate 
vectors of the box in which the coherent structures 
are located. In particular, if the system of coordinates consists of the vectors 
${\bf a}=(a_x,a_y)$
and ${\bf b}=(b_x,b_y)$, then the metric tensor becomes (in 2D)
$G=h^T h$, where 
$h$ is the coordinate 
transformation matrix with ${\bf a}$, ${\bf b}$ as its rows
[$(x_m,y_m)^T = h\cdot (\xi_n,\eta_n)^T$]. Then the Lagrangian for
the fictitious coupled dynamics of the ``particle''-lattice and the MD
box reads\cite{Rahman-Parrinello:a,Rahman-Parrinello:b}: 
\begin{equation} 
\label{LagrangianMD}
\begin{array} {rcl}
\ds {\cal L} &=&\ds  \frac{1}{2} \sum_n M (G_{11} \dot{\xi}_n^2 + 2 G_{12} \dot{\xi}_n
\dot{\nu}_n + G_{22} \dot{\nu}_n^2 )  \\[3.0ex]
&&\ds + \frac{1}{2} W (\dot{a}_x^2+\dot{a}_y^2+
\dot{b}_x^2 + \dot{b}_y^2)
-\sum_{m,n} V_{mn}.
\end{array} 
\end{equation}
where $M,W$ are the particle and box mass respectively, $(\xi_n,\nu_n)$ are
the coordinates of the $n$-th particle in the ``box frame'', $V_{mn}$ is the
interaction potential between quasi-particles $m$ and $n$, and the
summations are over the total number of vortices $N_v$.
Then, one can perform PR-MD by solving
the ensuing dynamical equations\cite{Rahman-Parrinello:a,Rahman-Parrinello:b},
to obtain the coordinate system evolution (along with that of the ``particle'' positions). 
More specifically, from (\ref{LagrangianMD}) one obtains the dynamics for each
of the  $4 N_v + 8$ degrees of freedom, say $q_n$ and $\dot q_n$, 
via the Euler-Lagrange equations
$d/dt(\partial {\cal L}/\partial \dot{q}_n) = \partial {\cal L}/\partial {q}_n$.
In order to emulate the thermodynamic limit of the system
(large number of vortices), the {\em minimal
image convention} can be employed where the box is periodically repeated
for all of its 8 neighbors and, for each vortex, the $N_v$
largest contributions, from all the 9 boxes, are taken into account.
%
%
The results obtained in Ref.\ [\refcite{rcg:30:vortexMD}] indicate that,
interacting spiral vortices in a Ginzburg-Landau field-theoretic 
context\cite{Aranson} (which, in fact, is a dissipative version of the GP 
equation---see also Ref.\ [\refcite{disgp}]),
typically settle, for sufficiently large number of vortices, to
configurations similar to the experimentally observed triangular configuration 
(see Fig.\ \ref{pfig8}). 
Note that similar results can be
applied to pulses where the topological charge is zero.

In order to specifically apply the PR-MD simulation to a gas of vortices in the
BEC model (\ref{peq1}), one needs to recast the MD Lagrangian
(\ref{LagrangianMD}) using the realistic vortex-vortex Lagrangian 
(\ref{LagrangianMD2})\cite{InProgress}.
The form of the Lagrangian (\ref{LagrangianMD}) allows for a direct 
incorporation of the effects induced by the MT and OL.
It is also possible to study the configuration changes
in the presence of quadrupolar excitations or other symmetry  
stresses exerted on the vortex 
lattice (motivated by the experiments of Refs.\ [\refcite{latt3,engel1,engel2}]).
We should note that the external imposition, in the box containing 
the vortices, of shear stresses of different symmetry is quite feasible
experimentally. It can be implemented through a rotating laser
manipulating the boundary of the box
(in this case, the periphery of the Thomas-Fermi cloud)\cite{engel3}.

\begin{figure}[ht]
\centerline{
\epsfig{file=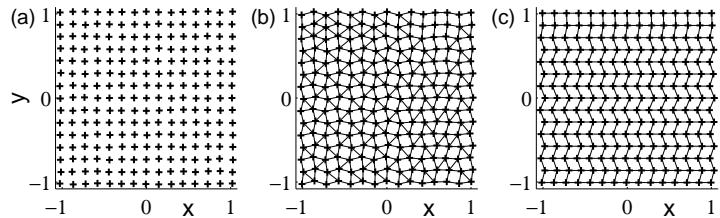, width=3.8in ,angle=0,silent=}
}
\caption[]{
Molecular dynamics simulations with an initial pattern (a) corresponding to a slightly
perturbed square lattice of $N_v=225$ same-charge vortices.
(b) After some transient, the configuration equilibrates to a
star pattern. (c) Rhomboidal pattern obtained from a 
slightly different initial perturbation of 
(a). 
\label{pfig81}
}
\end{figure}

The PR-MD approach is therefore a powerful tool that can be used 
to identify the conditions for which
the triangular state persists as the ground state configuration
(for such conditions, see for example Ref.\ [\refcite{Komineas:04}]). 
It can also be used to obtain the structural transition points
(obtained through direct simulations, e.g., in the
presence of the optical lattice in Ref.\ [\refcite{Pu:04}]) 
to other states such as rhomboidal or star-like patterns 
(see for example Fig.\ \ref{pfig81}). 
The approach consists of analyzing the dynamically
obtained PR-MD states through stability computations, 
standard continuation
and bifurcation theory tools to explore the structural phase transitions
and to obtain the instability eigenvectors that lead to phase
transformations. 
An alternative dynamical scenario, very relevant to recent experimental
settings, involves the annihilation of a central chunk of vortex lattice 
matter, through a localized laser heating\cite{engel2}, that
results in the remaining vortex lattice exhibiting oscillatory 
modes, known as Tkachenko oscillations.
Such modes may also be identified\cite{InProgress} as
limit-cycle, time-periodic solutions of the PR-MD numerical procedure.

\section{Summary and outlook}\label{sec:Conclusions}


While there is a large volume of work on the waves of topological charge
in Bose-Einstein condensates (a large fraction of which is summarized
			      in Ref.\ [\refcite{fetter}], as well as in
			      the present brief review), there are 
still numerous open problems regarding the vortex state that this
experimentally and theoretically tractable context may allow us to 
explore. 

  Clearly, a prominent position among such open problems 
(in the context of isolated vortices) is the
question of a mathematical understanding of the detailed dynamical
stability picture of vortices of various topological charges
  ($S\geq 1$) in the presence of the combined magnetic and optical
trappings. A first step in that direction is offered by the work
  of Ref.\ [\refcite{Pego}]; however there are still many open
questions concerning the effect of the trapping potentials.

Another subject that apparently has received very little attention
and whose theoretical understanding is still to a large extent
incomplete (in the context of few vortices) is 
the behavior of vortex dipoles. Such
configurations have been now obtained for two-component BECs
  in the work of Refs.\ [\refcite{bernard,Ueda:04}], however
topics such as the interaction of such dipoles and the ensuing
dynamics are still unexplored. From the mathematical point of
  view, a first attempt at obtaining reduced equations that
adequately describe dipole dynamics has been given in
  Ref.\ [\refcite{pknewton}]. The applicability of such an approach
in the context of BECs would be of particular interest.

There are also numerous interesting open questions regarding
vortex lattices. The dynamical imposition/time dependence 
of external potentials in conjunction with possible interaction
between multiple condensate components may produce interesting 
effects of frustration that may induce structural phase transitions
  such as the ones discussed in Ref.\ [\refcite{Pu:04}]
(e.g., a reshaping from a triangular lattice 
to a square one; see for example Ref.\ [\refcite{116}])
The PR-MD setup is straightforwardly amenable
to the inclusion of such external effects. By properly incorporating a
2D OL, we expect to induce a locally attractive 
(or locally repulsive) energy landscape
for each vortex at any required location.
This should enable us to {\em engineer} a 
rich variety of ``target'' lattice configurations
---provided that their energy is not far from a local minimum.
Applications of this (as well as statistical mechanics) techniques
to understand the ground state of vortex lattices under external
perturbations or multi-component interactions is bound
to provide interesting conclusions for the non-equilibrium
thermodynamics of such multi-vortex patterns. Notice that some
of these tasks (such as identifying stationary vortex lattice
states and computing the corresponding eigenvalues of linearization
around them) can be performed by methods that have been developed
in matrix-free numerical linear algebra\cite{117}. 
This can be done 
by using appropriately initialized short bursts of time evolution simulations
instead of the very expensive large Jacobian eigenvalue computations.

These are only some among the many questions/topics that are now 
starting to be addressed (for instance, one can ask the
same questions at finite temperature and try to understand
the interaction of the vortex condensate with the gas in
			  that case). We hope that 
this intriguing journey still hides, as Cavafy says in
his {\it Ithaca}, a lot of 
``ports seen for the first time'' \dots


%

\vspace{5mm}

{\bf Acknowledgments}

It is a real pleasure to acknowledge the invaluable contribution of 
our collaborators in many of the topics discussed here: Egor Alfimov, 
Alan Bishop, Fotis Diakonos, Todd Kapitula, 
Yuri Kivshar, Volodya Konotop, Boris Malomed, Dimitri Maroudas, Alex Nicolin,
Hector Nistazakis, Dmitry Pelinovsky, Mason Porter,
Nick Proukakis, Mario Salerno, Peter Schmelcher, Augusto Smerzi, Giorgos Theocharis 
and Andrea Trombettoni are greatly thanked. 
We would also like to acknowledge numerous 
useful discussions with Peter Engels and David Hall.
PGK is grateful to the Eppley Foundation for Research, the 
NSF-DMS-0204585 and the NSF-CAREER program for financial support.
RCG acknowledges the support of a SDSU Foundation Grant-In-Aid. 
DJF acknowledges the support of the Special Research Account of University of Athens.
IGK acknowledges the support of a NSF-ITR grant.

\end{document}